\documentclass[twocolumn, trackchanges]{aastex701}
\usepackage{booktabs, amsmath, gensymb, wasysym, dcolumn}

\begin{document}

\title{Multiple Components and Spectral Evolution of BL Lacertae as Revealed by Multiwavelength Variability and SED Modeling}

\author[orcid=0009-0007-1759-1803]{Hanxiao Xia}
\affiliation{School of Physics and Astronomy, Beijing Normal University, Beijing 100875, China}
\affiliation{Institute for Frontiers in Astronomy and Astrophysics, Beijing Normal University, Beijing 102206, China}
\email{202321160018@mail.bnu.edu.cn}

\author[orcid=0009-0006-9824-2498]{Ziming Wang} 
\affiliation{School of Physics and Astronomy, Beijing Normal University, Beijing 100875, China}
\affiliation{Institute for Frontiers in Astronomy and Astrophysics, Beijing Normal University, Beijing 102206, China}
\email{zmwang489@mail.bnu.edu.cn}

\author[orcid=0000-0002-8709-6759]{Jianghua Wu$^\dagger$}
\affiliation{School of Physics and Astronomy, Beijing Normal University, Beijing 100875, China}
\affiliation{Institute for Frontiers in Astronomy and Astrophysics, Beijing Normal University, Beijing 102206, China}
\email[show]{jhwu@bnu.edu.cn}
\correspondingauthor{Jianghua Wu}

\author[orcid=0000-0003-1878-9428]{Yue Fang}
\affiliation{College of Intelligent Systems Science and Engineering, Hubei Minzu University, Enshi 445000, China}
\email{fangyue@hbmzu.edu.cn}

\author[orcid=0009-0004-0627-7139]{Shiyu Du}
\affiliation{School of Physics and Astronomy, Beijing Normal University, Beijing 100875, China}
\affiliation{Institute for Frontiers in Astronomy and Astrophysics, Beijing Normal University, Beijing 102206, China}
\email{202421101125@mail.bnu.edu.cn}

\begin{abstract}

BL Lac has entered an active state since 2020, with multiwavelength observations revealing intense flares. In this study, we conducted 12-night multicolor optical monitoring using an 85 cm telescope from 2020 September to 2024 June and collected long-term broad-band archived data from radio to $\gamma$-rays. Intraday variabilities were detected on four nights, and most of them exhibited a bluer-when-brighter behavior. Both clockwise and counterclockwise spectral hysteresis loops were found within a single night. However, no reliable intraband time lag was detected for the intranight variabilities. On long timescales, the cross-correlation analysis shows that the variations of the optical, X-ray, and $\gamma$-ray bands do not reveal an obvious time delay, while the variations in the radio bands lagged them by about 370 days. The measured time lags suggest two distinct emission regions respectively responsible for the optical to $\gamma$-ray radiation and for the radio radiation, with a spatial separation of approximately $4.50\times10^{19}\ \rm cm$. We modeled the broad-band spectral energy distributions during four flaring epochs and one quiescent epoch, and found evidence for the possible persistent existence of a very high energy emission region. We also confirmed a spectral evolution of the source from an intermediate synchrotron peaked BL Lac object to a low synchrotron peaked BL Lac object.

\end{abstract}


\section{Introduction} \label{1}

Blazars are a peculiar subset of radio-loud active galactic nuclei (AGNs), whose most remarkable geometric feature is the presence of relativistic jets orienting at a small angle to our line of sight \citep{1995PASP..107..803U}. Distinguished by the strength of emission lines in the optical spectrum, blazars are further classified into BL Lac objects and flat-spectrum radio quasars (FSRQs). BL Lac objects are characterized by weak or absent emission features, with typical equivalent widths $\mathrm{EW} \leq 5\ \mathrm{\AA}$ \citep{1995PASP..107..803U}, while FSRQs exhibit stronger and quasar-like emission lines, indicative of a more luminous broad-line region (BLR). The alignment of jets with our line of sight causes strong Doppler-boosted nonthermal emission with high polarization, accompanied by rapid and large-amplitude variability across the entire electromagnetic spectrum and over diverse timescales \citep{1997ARA&A..35..445U}. 

In terms of the timescale of variations, long-term variability (LTV) spans from months to years, short-term variability (STV) lasts from days to months, and intraday variability (IDV) occurs on timescales as short as minutes to hours \citep{1995ARA&A..33..163W, 2012MNRAS.420.3147G, 2019ApJ...871..192P, 2024A&A...686A.228O}. Many blazars display a blending of the three variability patterns in the light curves. Along with flux variability, color/spectral behavior serves as a powerful tool for exploring the radiative properties of blazars. Three classes of color behaviors have been detected: bluer-when-brighter (BWB), redder-when-brighter (RWB), and achromatic trends. Some studies suggest that BL Lac objects tend to exhibit a BWB trend, while FSRQs show an RWB trend \citep{2006A&A...450...39G, 2012MNRAS.425.3002G, 2020MNRAS.492.1295P, 2022ApJ...926...91F, 2022ApJS..259...49Z}. In another important aspect, studies of time lags among multiwavelength variations help to probe the relative locations of emission regions and the propagation of disturbances in the jets \citep{2016A&A...593A..91A, 2023PASP..135i4101C, 2024A&A...692A.203K}. Sample analyses \citep[e.g.,][]{2014MNRAS.441.1899F, 2014MNRAS.445..428M, 2022MNRAS.510..469K} have revealed significant correlations and time lags of up to months between long-term trends in $\gamma$-ray and radio emissions for some sources.

The broad-band spectral energy distribution (SED) of a blazar usually shows a prominent double-humped structure in the logarithmic representation of $\nu F_{\nu}$ versus $\nu$ \citep{1998MNRAS.299..433F}. In the leptonic scenario, the low-energy component generated by relativistic electrons in the jet via synchrotron radiation ranges from radio to UV and even X-ray bands, whereas the high-energy component originating from inverse Compton (IC) scattering of low-energy photons by high-energy electrons covers X-ray and $\gamma$-ray bands, possibly reaching very high energy (VHE) $\gamma$-ray bands \citep{1996ApJ...461..657B, 2009ApJ...704...38S, 2012MNRAS.423..756P, 2020A&A...637A..86M}. When the seed photons are produced by synchrotron radiation itself in the jet and then up-scattered by energetic electrons, this process is called synchrotron self-Compton \citep[SSC,][]{2005A&A...440..845L, 2008ApJ...686..181F}. If the seed photons come from outside the jet (e.g., from the BLR, dusty torus, and accretion disc), the process is known as external Compton \citep[EC,][]{2009ApJ...692...32D, 2016ApJ...830...94F}. BL Lac objects are further classified according to the frequency of the synchrotron peak in their SEDs: those with $\nu_{\rm p}^{\rm sync}\leq 10^{14}\ \mathrm{Hz}$ are designated as low synchrotron peaked BL Lac (LBL) objects, those with $10^{14}< \nu_{\rm p}^{\rm sync}\leq 10^{15}\ \mathrm{Hz}$ as intermediate synchrotron peaked BL Lac (IBL) objects, and those with $\nu_{\rm p}^{\rm sync}> 10^{15}\ \mathrm{Hz}$ as high synchrotron peaked BL Lac (HBL) objects \citep{1995ApJ...444..567P}. A similar definition was expanded to all AGNs dominated by nonthermal emission: low-synchrotron-peaked blazars (LSP) for $\nu_{\rm p}^{\rm sync}\leq 10^{14}\ \mathrm{Hz}$, intermediate-synchrotron-peaked blazars (ISP) for $10^{14}< \nu_{\rm p}^{\rm sync}\leq 10^{15}\ \mathrm{Hz}$, and high-synchrotron-peaked blazars (HSP) for $\nu_{\rm p}^{\rm sync}> 10^{15}\ \mathrm{Hz}$ \citep{2010ApJ...716...30A, 2016ApJS..226...20F}.

BL Lac ($\mathrm{R.A.} = 22:02:43.29,\ \mathrm{decl.} = 42:16:39.98$, J2000) is the prototype of BL Lac objects, located at a redshift of $z = 0.0688 \pm 0.0002$ \citep{1977ApJ...212L..47M}. Owing to its broad-band activity and decades of observational campaigns, it has become one of the most well-investigated BL Lac objects \citep[e.g.,][]{2008Natur.452..966M, 2009A&A...507..769R, 2009A&A...501..455V, 2017MNRAS.469.3588M, 2022MNRAS.513.4645S}, and it is also the first blazar where IDV was observed \citep{1970ApJ...159L..99R, 1989Natur.337..627M}. Typically, BL Lac is often listed as an IBL \citep{2011ApJ...743..171A, 2016ApJS..226...20F, 2016A&A...592A..22H} or sometimes as an LBL \citep{2018A&A...620A.185N}, according to its synchrotron peak frequency. Moreover, since the detections of its VHE $\gamma$-ray photons have been reported in the literature \citep{2007ApJ...666L..17A, 2019A&A...623A.175M}, BL Lac is widely known as a TeV blazar.

It should be noted that BL Lac has entered an unprecedented active phase since 2020 January (MJD 58850), with multiple dramatic outbursts detected across all passbands. A maximum $\gamma$-ray flux of $(1.74\pm0.09)\times10^{-5}\ \mathrm{phs\ cm^{-2}\ s^{-1}}$ on MJD 59868.5 was reported in the daily binned light curve by \citet{2024MNRAS.527.5140S}. In X-ray energies, the source showed its brightest flare with a peak flux of $3.44\times10^{-10}\ \mathrm{erg\ cm^{-2}\ s^{-1}}$ in the 0.3--10 keV range on MJD 59128.17, and for the first time its spectral evolution from IBL to HBL was found during the high X-ray state \citep{2021MNRAS.507.5602P}. In contrast to the quasi-simultaneous flares in the $\gamma$-ray, X-ray, and optical bands, the radio enhancement was evidently delayed. \citet{2023ATel16340....1G} reported a flaring event with a flux density reaching 21.139 Jy on MJD 60262 in the millimeter waveband, more than 30\% higher than any previous flare of BL Lac in the past 20 years. Besides, brightening fluxes were discovered in 15 and 43 GHz with Very Long Baseline Array (VLBA) observation \citep{2025ApJ...978...43M}. 

For the post-2020 flaring phase, the systematic cross-correlation analysis of multiwavelength variations, as well as SED modeling associated with the radio flaring states, remain absent. In addition, the optical variability, on either intraday or longer timescales, of BL Lac during this active state deserves sustained investigation and discussion. Therefore, we performed intraday optical observations and collected publicly available multiwavelength data from MJD 54477 to 60760. We analyzed the variability behavior of the source on different timescales and modeled the SEDs in multi-epochs. Observations and data reduction for datasets are described in Section~\ref{2}. In Section~\ref{3}, we explore the IDV, color behavior, and time lags of the intraday observations. We analyze long-term multiwavelength variability and cross-correlations, and estimate the relative distances between radiation regions in Section~\ref{4}. In Section~\ref{5}, the results of SED modeling for multi-epochs are presented, revealing the existence of multiple emission components and differences in their physical parameters. The main conclusions of this work are summarized in Section~\ref{6}.

\section{Observations, Data Acquisition and Reduction} \label{2}

In this study, we performed intraday monitoring in optical bands and retrieved from public archives the multiwavelength data from radio to $\gamma$-ray energies for BL Lac. This section describes the observations and processing procedures for these various datasets.

\subsection{Optical Data: Intraday Observations} \label{2.1}

The intraday observational campaign was carried out with the 85 cm telescope at Xinglong Station of the National Astronomical Observatories, Chinese Academy of Sciences (NAOC).This telescope uses the prime-focus optical design with a focal ratio of $F/3.3$. The CCD is a $2048 \times 2048$ chip with a field of view (FOV) of $\sim 32.8 \times 32.8\ \mathrm{arcmin}^2$. We observed BL Lac quasi-simultaneously in the \emph{BVR} bands on 12 nights from 2020 September to 2024 June. A total of nearly 5,000 data points were collected, with the corresponding observational information summarized in Table~\ref{tab1}.

\begin{deluxetable*}{lccccccccccccc}[ht!]
\tablewidth{0pt}
\tabletypesize{\footnotesize}
\tablecaption{IDV Results of BL Lac \label{tab1}}
\tablehead{
\colhead{MJD} &\colhead{Date} & \colhead{Filter} & \colhead{$N$} & \colhead{Duration} & \colhead{SD} & \multicolumn{3}{c}{Power-enhanced $F$-test} & \multicolumn{3}{c}{ANOVA test} & \colhead{Var?} & \colhead{Amp}\\
\cmidrule(lr){7-9} \cmidrule(lr){10-12}
& & & & \colhead{(hr)} & & \colhead{$\nu_1/\nu_2$} & \colhead{$F$} & \colhead{$F_c$} & \colhead{$\nu_1/\nu_2$} & \colhead{$F$} & \colhead{$F_c$} & & \colhead{(\%)}\\
(1) & (2) & (3) & (4) & (5) & (6) & (7) & (8) & (9) & (10) & (11) & (12) & (13) & (14)
}
\startdata
& & $B$ & 170 & 4.44 & 0.011 & 169/507 & 12.09 & 1.33 & 33/136 & 20.60 & 1.81 & V & 12.40 $\pm$ 0.41 \\
59110 & 20200918 & $V$ & 170 & 4.44 & 0.008 & 169/507 & 14.42 & 1.33 & 33/136 & 30.64 & 1.81 & V & 10.23 $\pm$ 0.25 \\
& & $R$ & 170 & 4.44 & 0.008 & 169/507 & 14.10 & 1.33 & 33/136 & 21.41 & 1.81 & V & 9.63 $\pm$ 0.21 \\
\hline
& & $B$ & 200 & 5.23 & 0.009 & 199/597 & 29.95 & 1.30 & 39/160 & 41.29 & 1.73 & V & 15.95 $\pm$ 0.37 \\
59111 & 20200919 & $V$ & 197 & 5.23 & 0.006 & 196/588 & 63.75 & 1.30 & 39/157 & 54.26 & 1.73 & V & 15.18 $\pm$ 0.23 \\
& & $R$ & 199 & 5.23 & 0.006 & 198/594 & 56.54 & 1.30 & 39/159 & 56.58 & 1.73 & V & 13.78 $\pm$ 0.20 \\
\hline
& & $B$ & 20 & 0.50 & 0.031 & 19/57 & 0.28 & 2.24 & 3/16 & 1.28 & 5.29 & N & \\
59112 & 20200920 & $V$ & 20 & 0.50 & 0.009 & 19/57 & 1.63 & 2.24 & 3/16 & 1.30 & 5.29 & N & \\
& & $R$ & 20 & 0.50 & 0.011 & 19/57 & 1.06 & 2.24 & 3/16 & 3.08 & 5.29 & N & \\
\hline
& & $B$ & 8 & 0.55 & 0.012 & 7/21 & 0.94 & 3.64 & 1/6 & 1.97 & 13.75 & N & \\
59113 & 20200921 & $V$ & 8 & 0.55 & 0.006 & 7/21 & 2.14 & 3.64 & 1/6 & 6.69 & 13.75 & N & \\
& & $R$ & 8 & 0.55 & 0.004 & 7/21 & 3.62 & 3.64 & 1/6 & 2.66 & 13.75 & N & \\
\hline
& & $B$ & 161 & 2.15 & 0.032 & 160/480 & 1.13 & 1.34 & 32/128 & 1.21 & 1.83 & N & \\
59565 & 20211217 & $V$ & 160 & 2.14 & 0.022 & 159/477 & 1.96 & 1.34 & 31/128 & 1.64 & 1.84 & N & \\
& & $R$ & 160 & 2.14 & 0.021 & 159/477 & 1.49 & 1.34 & 31/128 & 1.65 & 1.84 & N & \\
\hline
& & $B$ & 200 & 2.68 & 0.030 & 199/597 & 0.63 & 1.30 & 39/160 & 0.21 & 1.73 & N & \\
59566 & 20211218 & $V$ & 200 & 2.68 & 0.024 & 199/597 & 0.81 & 1.30 & 39/160 & 0.19 & 1.73 & N & \\
& & $R$ & 200 & 2.68 & 0.024 & 199/597 & 0.57 & 1.30 & 39/160 & 0.46 & 1.73 & N & \\
\hline
& & $B$ & 163 & 2.18 & 0.035 & 162/486 & 0.80 & 1.34 & 32/130 & 0.20 & 1.82 & N & \\
59567 & 20211219 & $V$ & 163 & 2.18 & 0.027 & 162/486 & 0.85 & 1.34 & 32/130 & 0.38 & 1.82 & N & \\
& & $R$ & 163 & 2.18 & 0.028 & 162/486 & 0.88 & 1.34 & 32/130 & 0.47 & 1.82 & N & \\
\hline
& & $B$ & 199 & 2.96 & 0.032 & 198/594 & 1.15 & 1.30 & 39/159 & 1.25 & 1.73 & N & \\
59568 & 20211220 & $V$ & 199 & 2.96 & 0.022 & 198/594 & 1.46 & 1.30 & 39/159 & 0.87 & 1.73 & N & \\
& & $R$ & 199 & 2.96 & 0.020 & 198/594 & 1.81 & 1.30 & 39/159 & 1.25 & 1.73 & N & \\
\hline
& & $B$ & 130 & 1.58 & 0.022 & 129/387 & 1.50 & 1.38 & 25/104 & 3.60 & 1.96 & V & 8.08 $\pm$ 1.01 \\
59887 & 20221104 & $V$ & 130 & 1.58 & 0.015 & 129/387 & 1.63 & 1.38 & 25/104 & 3.27 & 1.96 & V & 6.02 $\pm$ 0.59 \\
& & $R$  & 130 & 1.58 & 0.014 & 129/387 & 1.71 & 1.38 & 25/104 & 4.38 & 1.96 & V & 6.47 $\pm$ 0.45 \\
\hline
& & $B$ & 352 & 5.07 & 0.037 & 351/1053 & 2.39 & 1.22 & 70/281 & 8.75 & 1.52 & V & 18.55 $\pm$ 2.77 \\
59888 & 20221105 & $V$ & 352 & 5.07 & 0.022 & 351/1053 & 6.78 & 1.22 & 70/281 & 24.60 & 1.52 & V & 17.64 $\pm$ 1.74 \\
& & $R$ & 352 & 5.07 & 0.019 & 351/1053 & 7.60 & 1.22 & 70/281 & 37.82 & 1.52 & V & 15.64 $\pm$ 0.80 \\
\hline
& & $B$ & 38 & 2.28 & 0.008 & 37/111 & 1.36 & 1.80 & 7/30 & 1.46 & 3.30 & N & \\
60463 & 20240602 & $V$ & 38 & 2.28 & 0.004 & 37/111 & 1.22 & 1.80 & 7/30 & 0.41 & 3.30 & N & \\
& & $R$ & 38 & 2.28 & 0.005 & 37/111 & 1.37 & 1.80 & 7/30 & 0.43 & 3.30 & N & \\
\hline
& & $B$ & 25 & 1.87 & 0.012 & 24/72 & 2.03 & 2.06 & 4/20 & 1.42 & 4.43 & N & \\
60464 & 20240603 & $V$ & 25 & 1.87 & 0.005 & 24/72 & 1.37 & 2.06 & 4/20 & 0.26 & 4.43 & N & \\
& & $R$ & 25 & 1.87 & 0.006 & 24/72 & 1.26 & 2.06 & 4/20 & 1.37 & 4.43 & N & \\
\enddata
\tablecomments{The columns are (1) the Modified Julian Date, (2) the calendar date, (3) filter, (4) the number of exposures, (5) the total monitoring duration, (6) standard deviations of the differential magnitudes between the check star and the comparison star, (7–9) two degrees of freedom, $F$, and the critical value $F_c$ at the 99\% confidence level in the power-enhanced $F$-test, (10–12) two degrees of freedom, $F$, and $F_c$ at the 99\% confidence level in the ANOVA test, (13) variable or not (``V" is variable and ``N" is nonvariable), and (14) variability amplitude, respectively.}
\end{deluxetable*}

\begin{figure}[t!]
\begin{center}
    \includegraphics[angle=0,scale=0.32]{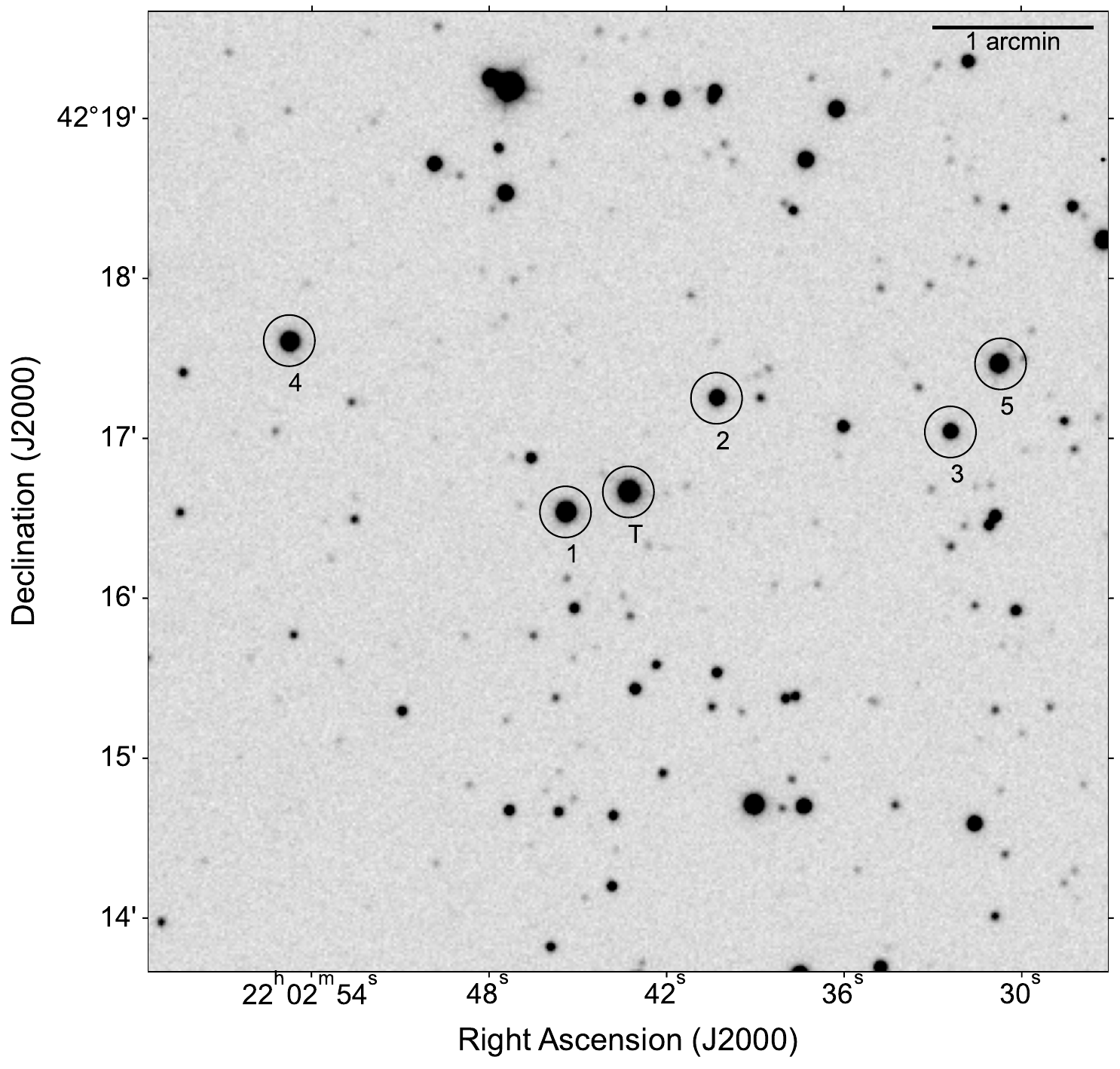}
    \caption{Finding chart of BL Lac in the $V$-band. The labels “T” and “1–5” represent BL Lac, two comparison stars, the check star, and two field stars, respectively.
    \label{fig1}}
\end{center}
\end{figure}

\begin{figure*}[ht!]
\plotone{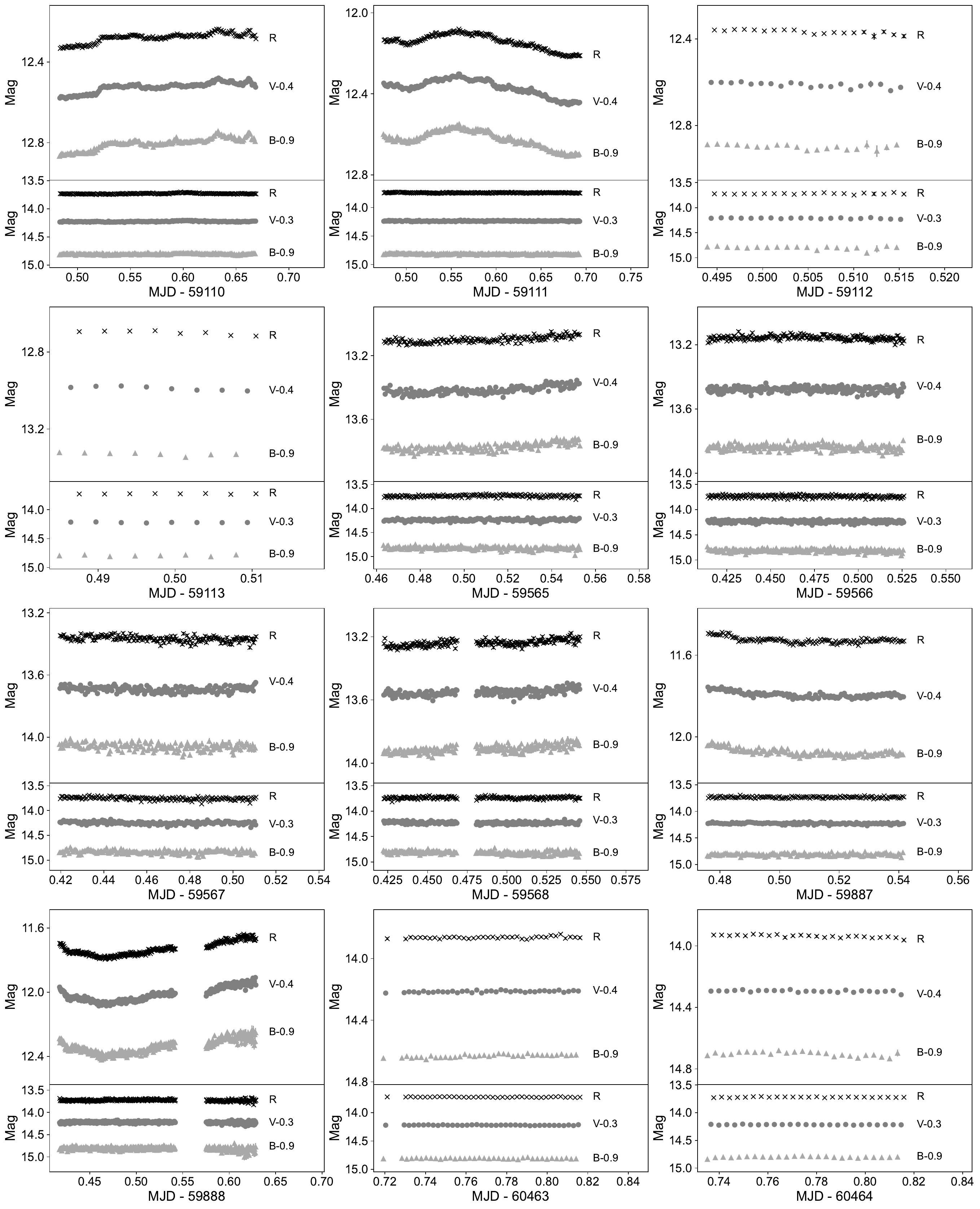}
\caption{The intraday light curves of BL Lac and the check star are displayed in the top and bottom panels of each subfigure, respectively. The $B$- and $V$-band light curves are shifted for clarity. On MJDs 59568 and 59888, the monitoring interruptions occurred temporarily due to bad weather. From MJD 59565 to 59568 and on MJD 59888, unfavorable weather conditions led to a relative large standard deviation for the magnitude of the check star, as shown in Table~\ref{tab1}.
\label{fig2}}
\end{figure*}

BL Lac and five stars are shown in Figure~\ref{fig1}, where stars 1 and 2 serve as comparison stars, star 3 is used as the check star, and stars 4 and 5 are field stars. The raw data were reduced following the standard procedure using the Python package \texttt{Photutils} version 1.12.0\footnote{\url{https://photutils.readthedocs.io/en/stable/}}\citep{2024zndo..10967176B}, including bias subtraction and flat-fielding. We used the \texttt{Background2D} function to estimate the background flux, and extracted instrumental magnitudes using the \texttt{aperture\_photometry} function. To mitigate the potential influence of random uncertainty associated with a single comparison star, the instrumental magnitudes of stars 1 and 2 were combined to produce the magnitudes of a synthetic comparison star with improved robustness. To determine the optimal aperture, photometry was performed using 10 different aperture radii ranging from 1 to 3 times the FWHM of stellar images. The final aperture radius was set to 1.8 times the FWHM, which produced the minimum standard deviation of the differential magnitudes between the check star and the comparison star (see Table~\ref{tab1}). As shown in Figure~\ref{fig2}, the light curves of BL Lac and the check star are presented. A few outliers resulting from poor weather or instrumental problems were excluded from the following analysis.

\subsection{Optical Data: Long-term Observations} \label{2.2}

To supplement the optical data observed by us, we retrieved and compiled long-term light curves from various public archives.
The American Association of Variable Star Observers\footnote{
\url{https://www.aavso.org}
}(AAVSO) maintains a database of almost 70 million variable star observations. The photometric data of BL Lac in the \emph{BVRI} bands were downloaded from the official website, covering the period from MJD 54690 to 60618. The magnitudes with only upper limits were manually removed.
The Zwicky Transient Facility\footnote{\url{https://irsa.ipac.caltech.edu/Missions/ztf.html}}\citep[ZTF,][]{2019PASP..131a8002B, 2019PASP..131a8003M} delivers optical imagery for time-domain astrophysics analysis at the Palomar Observatory. The $g$-band and $r$-band flux data, during the time range MJD 58234--60610, were collected. To ensure data quality, only data points with catflags = 0 were retained.
The All-Sky Automated Survey for Supernovae \citep[ASAS-SN,][]{2019MNRAS.485..961J} nightly monitors the entire visible sky using a global network of 24 telescopes. The $V$-band and $g$-band photometric data of the target were obtained from ASAS-SN Sky Patrol V2.0\footnote{\url{http://asas-sn.ifa.hawaii.edu/skypatrol/}}\citep{2014ApJ...788...48S, 2023arXiv230403791H}, encompassing the period between MJD 57007 and 60646. Observations flagged as ``bad" were discarded from the analysis.
The Katzman Automatic Imaging Telescope (KAIT)\footnote{\url{http://herculesii.astro.berkeley.edu/kait/agn/}} records the variability of a total of 163 AGNs. We acquired the light curve of BL Lac spanning from MJD 55808 to 60254. Although KAIT performs unfiltered photometry, its measurements correspond approximately to the $R$-band in practice \citep{2003PASP..115..844L}.

\subsection{$\gamma$-ray Data: Fermi-LAT} \label{2.3}

The Large Area Telescope on board the Fermi Gamma-ray Space Telescope (Fermi-LAT) surveys the entire sky every 3 hours, covering an energy range from below 20 MeV to above 300 GeV\citep{2009ApJ...697.1071A}.

For SED modeling, the publicly available data of BL Lac (4FGL J2202.7+4216) in the energy range of 0.1--300 GeV were collected from the Fermi-LAT data server\footnote{\url{https://fermi.gsfc.nasa.gov/ssc/data/access/}}. The unbinned likelihood analysis was performed using version 2.4.0 of the Fermi Science Tools \citep{2019ascl.soft05011F}. We utilized analysis cuts of a circular region of interest (ROI) of 15\degree, evclass = 128, evtype = 3, and zmax = 90 of the photon data. The XML model files were generated using the \texttt{make4FGLxml.py} script, with the isotropic background model \texttt{iso\_P8R3\_SOURCE\_V3\_v1.txt}, and the Galactic diffuse emission model \texttt{gll\_iem\_v07.fits}. The parameters of other sources within 5\degree\ of the ROI center were allowed to vary, whereas those for sources between 5\degree\ and 15\degree\ were fixed to their values in the 4FGL catalogue \citep{2022ApJS..260...53A, 2023arXiv230712546B}. The $\gamma$-ray spectrum of BL Lac was described by a log-parabola function in the template. The criterion for a significant detection was TS $>$ 10. 

In addition, the 0.1--100 GeV 3 day binned light curve of BL Lac, spanning from MJD 54683 to 60760, was obtained from the Fermi-LAT Lightcurve Repository \citep[LCR,][]{2023ApJS..265...31A}\footnote{\url{https://fermi.gsfc.nasa.gov/ssc/data/access/lat/LightCurveRepository/}}, and those measurements with upper limits were excluded.

\subsection{X-ray Data: Swift-XRT} \label{2.4}

The Swift X-ray Telescope \citep[XRT,][]{2005SSRv..120..165B} observes in the 0.3--10 keV energy range, and has accumulated long-term data of BL Lac over the past years. 
We retrieved the 0.3--10 keV light curve of BL Lac using the online Swift-XRT data product generator tool\footnote{\url{https://www.swift.ac.uk/user_objects/}} provided by the UK Swift Science Data Center \citep[SDC,][]{2007A&A...469..379E, 2009MNRAS.397.1177E}, which includes all observations in both the photon counting (PC) mode and the windowed timing (WT) mode. 

For the SED fitting, the PC mode event files (level 2) of desired dates were downloaded from the SDC. 
The spectrum on each date was extracted using \texttt{xrtproduct} version 0.4.3 (\texttt{HEASoft} version 6.35.1d) and the XRT CALDB release 20240522. 
The source extraction region was set to an annulus with an outer bound of 40 pixels. 
The inner exclusion radii (from 0 to 3 pixels) were set separately for each observation to avoid the pile-up effect, with reference to the SDC pile-up thread\footnote{\url{https://www.swift.ac.uk/analysis/xrt/pileup.php}}. 
The background region was set to an annulus with inner and outer radii of 80 and 120 pixels, respectively, and was centered on the source position. 
Energy channels were grouped using the \texttt{grppha} tool requiring a minimum of 20 photons per bin.  
Spectra within 0.4--10 keV were fitted by XSPEC version 12.14.0b. 
The absorption model \texttt{tbabs} \citep[][]{2000ApJ...542..914W} was adopted to account for the galactic contribution, with $N_{\rm H}$ value of $0.29 \times 10^{22}\ \mathrm{cm^{-2}}$ obtained from SDC nH tool \citep[][]{2013MNRAS.431..394W}. 
Fit results of absorbed power-law model (\texttt{tbabs*powerlaw}) were used to model the SED in Section~\ref{5}. 

\subsection{Radio Data: SMA and VLBA} \label{2.5}

The Submillimeter Array \citep[SMA,][]{2004ApJ...616L...1H} is an eight-element radio interferometer located on Maunakea in Hawaii, operating at frequencies between 180 and 418 GHz. The SMA calibrator list\footnote{\url{http://sma1.sma.hawaii.edu/callist/callist.html}} frequently obtains dedicated observations of $\sim$ 1900 sources at 1 mm and 850 $\mu$m \citep{2007ASPC..375..234G}. The long timescale light curve in the 1 mm band of BL Lac was obtained from the SMA calibrator list, covering the period from MJD 54477 to 60673.

Nearly simultaneously, a significant flux increase of BL Lac was also observed in the centimeter bands. Peak fluxes of 12.336 Jy (MJD 60407) at 15 GHz and 16.278 Jy (MJD 60273) at 43 GHz were observed by the VLBA MOJAVE program\footnote{\url{https://www.cv.nrao.edu/MOJAVE/allsources.html}}\citep{2018ApJS..234...12L} and BEAM-ME program\footnote{\url{https://www.bu.edu/blazars/BEAM-ME.html}}\citep{2016Galax...4...47J}, respectively. The light curves from both programs were utilized for subsequent scientific analysis.

\begin{figure*}[ht!]
\plotone{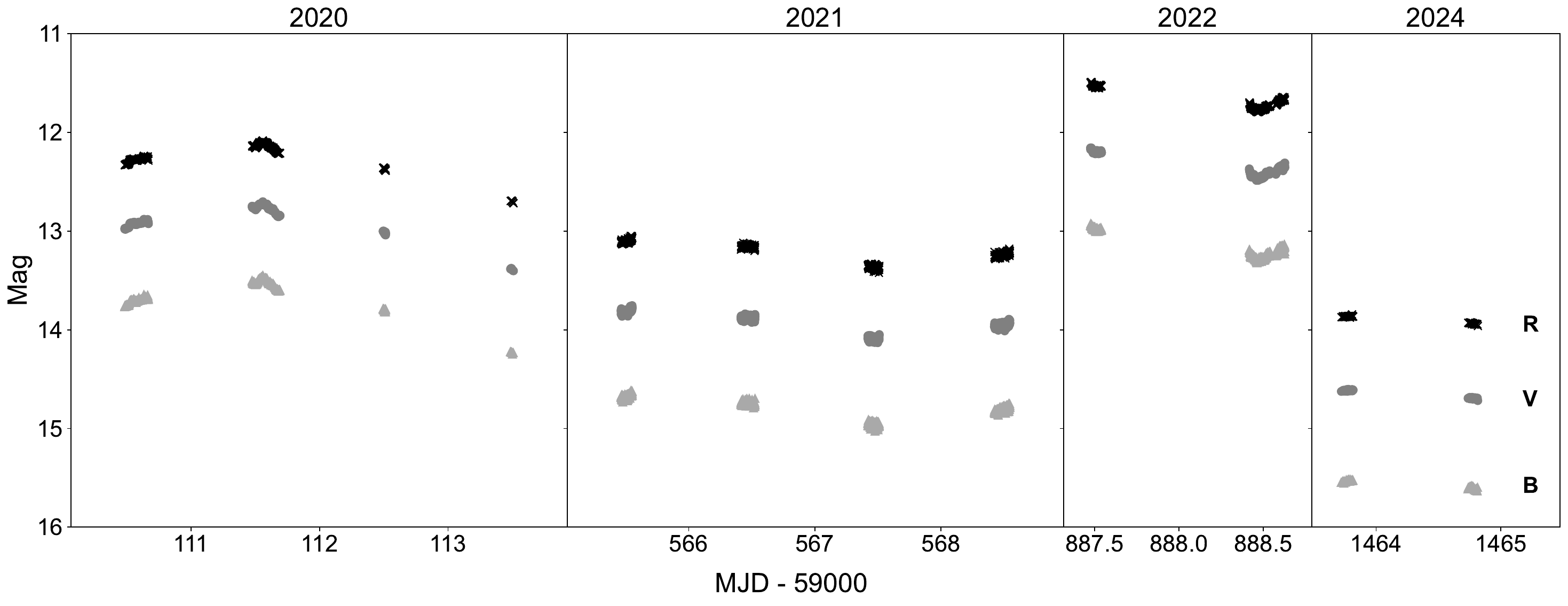}
\caption{The overall $B$-, $V$-, and $R$-band light curves of BL Lac.
\label{fig3}}
\end{figure*}

\section{Short-term Variability Analysis} \label{3}

\subsection{Light Curves} \label{3.1}

Figure~\ref{fig3} presents the overall light curves of BL Lac from the 12 intraday observations. Figures~\ref{fig2} and~\ref{fig3} demonstrate that BL Lac exhibits intraday and inter-day optical variability, with a high degree of correlations among the $B$-, $V$-, and $R$-bands.

The source reached its brightest state on MJD 59887 with a maximum magnitude of 12.17 mag in the $V$-band, and its faintest state on MJD 60464 with 14.72 mag, yielding a total variability amplitude of 2.55 mag, indicating significant optical variability during this period. Through preliminary visual inspection, considerable IDVs are found on MJDs 59110, 59111, and 59888, while marginal IDVs are exhibited on MJDs 59565, 59568, and 59887.

\subsection{IDV Detection} \label{3.2}

To quantitatively analyze the IDV of BL Lac, we employed two statistical tests: the power-enhanced version of the $F$-test \citep{2014AJ....148...93D} and the one-way analysis of variance \citep[ANOVA,][]{1998ApJ...501...69D}. The power-enhanced $F$-test assesses the presence of IDV by the variance ratio between the light curves of the source and the check stars, while ANOVA evaluates whether IDV exists by dividing an individual light curve into groups and comparing the variance between groups and the whole light curve. Both methods have been widely adopted in previous variability studies \citep{2012MNRAS.425.3002G, 2017MNRAS.469.3588M, 2020ApJ...900..137W, 2022ApJ...933..224F, 2023ApJS..269...60Y}.

The results of the IDV detection are listed in Table~\ref{tab1}. $\nu$ represents the degree of freedom in both statistical tests. For the power-enhanced $F$-test, it is calculated by $\nu_1=N_{\rm b}-1$ and $\nu_2=N_{\rm c}-k$, where $N_{\rm b}$ is the number of exposures of BL Lac, $N_{\rm c}$ is the total number of exposures of check stars, and $k$ is the number of check stars (in our case, stars 3--5 were used, thus $k=3$). For ANOVA, it is calculated by $\nu_1=g-1$ and $\nu_2=N_{\rm b}-g$, where $g$ is the number of groups.  If the $F$ value of a light curve exceeds the critical value $F_c$, at the 99\% confidence level, the null hypothesis will be rejected, indicating the detection of variability. BL Lac exists variability (V) only when both tests are passed; otherwise, it is deemed nonvariable (N). The statistical methods confirm the presence of IDV on MJDs 59110, 59111, 59887, and 59888. No IDV was detected on the remaining eight nights. The $V$- and $R$-band light curves on MJDs 59565 and 59568 passed the power-enhanced $F$-test but failed the ANOVA test, and neither test detected IDV in the $B$-band light curves. Considering that the three bands are close in the frequency, it is unreasonable that variability would appear in part of them. Hence, we conclude that IDV was not present in the three bands on those two nights. The discrepancy between visual inspection and statistical methods on MJDs 59565 and 59568 may be due to the variation amplitude being too small and the relatively poor photometric precision caused by suboptimal weather, as also discussed in earlier works \citep{2021RAA....21..259L, 2022ApJ...926...91F}.

The variation amplitudes of those light curves labeled ``V" were computed according to the formula proposed by \citet{1996A&A...305...42H}:
\begin{equation} \label{eq1}
\mathrm{Amp} = 100\%\times\sqrt{(A_{\mathrm{max}}-A_{\mathrm{max}})^2 - 2 \sigma^2},
\end{equation}
where $A_\mathrm{max}$ and $A_\mathrm{min}$ are the maximum and minimum magnitudes, respectively, and $\sigma$ is the measurement error. Following \citet{2022ApJ...933..224F}, the uncertainty of $\mathrm{Amp}$ can be given by 
\begin{equation} \label{eq2}
\sigma_\mathrm{Amp}=(\frac{A_\mathrm{max}-A_\mathrm{min}}{\mathrm{Amp}})\times\sqrt{(\sigma_{A_\mathrm{max}}^2+\sigma_{A_\mathrm{min}}^2)}, 
\end{equation}
where $\sigma_{A_\mathrm{max}}$ and $\sigma_{A_\mathrm{min}}$ are the measurement errors corresponding to $A_\mathrm{max}$ and $A_\mathrm{min}$, respectively. The amplitudes and their associated uncertainties are also given in Table~\ref{tab1}. The largest amplitude appears in the $B$-band light curve on MJD 59888, reaching $18.55\pm2.77\%$. \citet{2023ApJS..269...60Y} and \citet{2025MNRAS.537.2586A} found even higher IDV amplitudes in the $B$- and $g$-bands. Except for the case on MJD 59887 where the amplitude in the $R$-band marginally surpasses (by less than 0.5\%) that in the $V$-band, all other instances show an increasing trend of amplitude with higher energy band, suggesting spectral evolution of the source. As reported in \citet{2021MNRAS.507.5602P}, the optical--UV spectrum of BL Lac tends to be flatter at brighter states and steeper at fainter ones, indicating a shift of the synchrotron peak to higher energy in high-flux states. In general, our findings are consistent with those reported in the literature \citep{1998AJ....115.2244W, 2017MNRAS.469.3588M, 2022ApJ...926...91F}.

\subsection{Color Behavior\label{3.3}}

\begin{figure*}[ht!]
\plotone{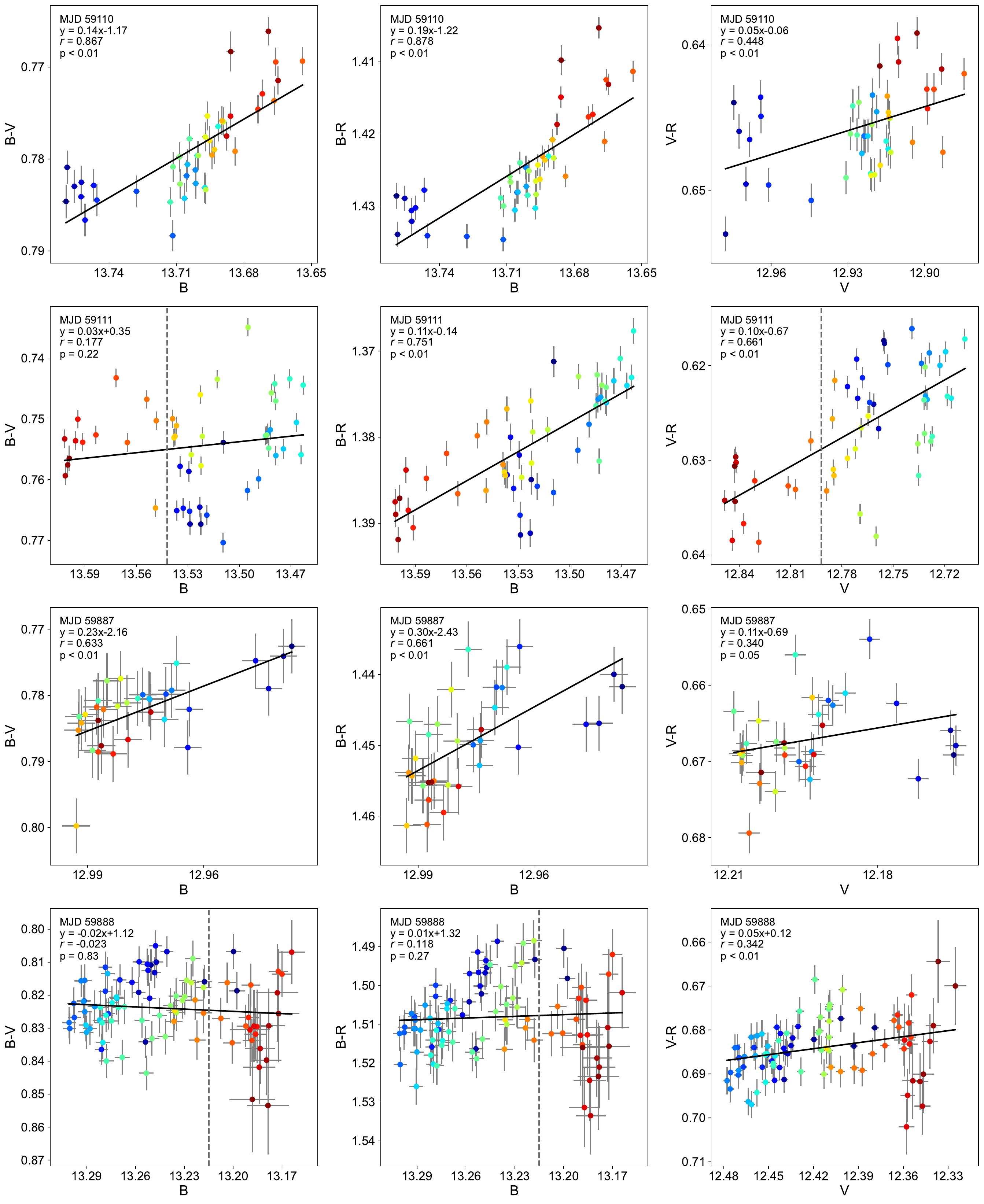}
\caption{CMDs of BL Lac for the nights exhibiting IDV. The color gradient from blue to red indicates the progression of observation time. The solid lines indicate the linear fitting results. The observation date (MJD), the linear regression equation, the correlation coefficient $r$, and the $p$-value are shown in the upper-left corner. The vertical dashed lines in the four subplots mark the data segments associated with spectral hysteresis loops.
\label{fig4}}
\end{figure*}

For the nights at which IDV was detected, we plotted color–magnitude diagrams (CMDs) to investigate the color behavior of BL Lac. The data points were binned at 5-min interval, and the color indices (CIs) of $B-V$, $B-R$, and $V-R$ were calculated. We employed the bivariate correlated errors and intrinsic scatter \citep[BCES,][]{2012Sci...338.1445N}\footnote{\url{https://github.com/rsnemmen/BCES}} method to perform the linear regression between CIs and magnitudes, which accounts for measurement errors in both variables \citep{1996ApJ...470..706A}. The significance of the correlation was then evaluated using the Spearman correlation coefficient. A reliable color–magnitude correlation was considered to exist only when the absolute value of the coefficient $r$ exceeded 0.2 and the level of statistical significance was greater than 99\% (that is, $p<0.01$). The results are displayed in Figure~\ref{fig4}.

Based on the above criteria, BL Lac exhibits diverse color behaviors. In eight subplots of Figure~\ref{fig4}, the source shows BWB trends, which is a common phenomenon in BL Lac objects \citep[e.g.,][]{2015MNRAS.452.4263G, 2023ApJ...943..135K, 2023ApJS..265...51A, 2024MNRAS.528.6823L}, but in others no significant trends are observed. Furthermore, instances of color reversal have also been reported in some objects. On the long timescale, \citet{2020MNRAS.498.3578S} reported two blazars exhibiting color reversals: AO 0235+164 and 3C 279. \citet{2020ApJ...902...41W} found a transition in AO 0235+164 from a BWB trend to an RWB trend and eventually to an achromatism. On the intraday timescale, \citet{2022ApJ...926...91F} reported a color reversal in BL Lac. However, this behavior is not found in our study.

Some theories have been proposed to explain the color behavior of BL Lac objects. \citet{2023MNRAS.522..102R} and some references therein concluded that the long-term color behavior is nearly achromatic due to variations of the Doppler factor, whereas the short-term behavior tends to exhibit strong chromatic trends, which are mainly driven by intrinsic energetic processes. According to the shock-in-jet model developed in \citet{1998A&A...333..452K}, a BWB trend occurs in a band where the electron acceleration timescale is significantly shorter than the cooling timescale. When the electron injection rate increases, high-energy electrons responsible for the high-frequency synchrotron radiation are produced very rapidly, leading to a hardening of the injected electron energy distribution (EED). This results in the high-frequency flux rising faster than the low-frequency flux, causing a BWB trend in the band.

Interestingly, the CMDs of $B-V\ \mathrm{vs.}\ B$ and $V-R\ \mathrm{vs.}\ V$ on MJD 59111 show spectral hysteresis loops, as seen on the right sides of the dashed lines. The former traces a counterclockwise pattern, while the latter follows a clockwise path. This is the first time that two modes of spectral hysteresis have been identified simultaneously within one night in BL Lac. Prior to this work, only \citet{2022ApJ...926...91F} documented two clockwise loops in BL Lac. A comparison with the intraday light curve on MJD 59111 (see Figure~\ref{fig2}) suggests that this behavior corresponds to a minor flare during that night. In addition, similar counterclockwise loops are also present in the CMDs of $B-V\ \mathrm{vs.}\ B$ and $B-R\ \mathrm{vs.}\ B$ on MJD 59888, appearing on the left sides of the dashed lines. Spectral hysteresis has been reported in other blazars on both intraday and day-to-month timescales. For example, it has been observed in the X-ray spectral behaviors of PKS 2155--304 \citep{1993ApJ...404..112S, 2014MNRAS.444.1077K}, 1ES 1218+304 \citep{2025MNRAS.539.2806K}, and Mrk 421\citep{1996ApJ...470L..89T, 2024MNRAS.529.1450G}, as well as in optical behaviors of PG 1553+113 \citep{2021A&A...645A.137A}, OJ 287 \citep{2020MNRAS.492.1295P}, and S5 0716+714 \citep{2013ApJS..204...22D, 2016MNRAS.456.3168M}. Theoretically, the pattern of spectral hysteresis depends on the ratio between the electron acceleration and cooling timescales \citep{1998A&A...333..452K}. When the cooling timescale surpasses the acceleration timescale, a clockwise loop (i.e., a ``soft lag") is typically observed. Conversely, when the two timescales are comparable, a counterclockwise loop, or a ``hard lag", tends to arise. Moreover, the clockwise loop is the expected behavior for frequencies below the synchrotron peak frequency, and the counterclockwise loop occurs at those near the synchrotron peak frequency. This implies that during the minor flare detected on MJD 59111, the electrons responsible for lower-energy photon emission experienced a faster acceleration process, whereas for electrons producing higher-energy photons, the acceleration and cooling timescales were roughly equal. Observing both types of loops suggests that the synchrotron peak frequency was located near the $B$-band. In the $B–V$ bands, the timescales of two processes achieved balance, while in the $V–R$ bands, the acceleration process proceeded faster than the cooling process.

\subsection{Cross-correlation Analysis for IDV\label{3.4}}

\begin{deluxetable*}{lcccr@{}lcr@{}lc}[ht!]
\tablewidth{0pt}
\tabletypesize{\normalsize}
\tablecaption{Results of Cross-correlation Analysis for IDV\label{tab2}}
\tablehead{
\colhead{MJD} & \colhead{Bands} & \multicolumn{4}{c}{ICCF} & \multicolumn{3}{c}{\texttt{PyROA}} & \colhead{ZDCF} \\
\cmidrule(lr){3-6} \cmidrule(lr){7-9}
\colhead{} & \colhead{} & \colhead{$r_{\rm max}$} & \colhead{$f_{\rm out}$ (\%)} & \multicolumn{2}{c}{Lag (min)} & \colhead{$f_{\rm out}$ (\%)} & \multicolumn{2}{c}{Lag (min)} & \colhead{Lag (min)}
}
\startdata
 & $B-V$ & 0.98 & 0 & $-0.7$ & $^{+1.8}_{-1.7}$ & 0 & $-0.6$ & $^{+0.7}_{-0.7}$ & $0.6^{+0.7}_{-1.8}$ \\
59110 & $B-R$ & 0.97 & 0 & $-1.1$ & $^{+2.0}_{-1.7}$ & 0 & $-0.7$ & $^{+0.8}_{-0.8}$ & $1.1^{+0.7}_{-2.1}$ \\
 & $V-R$ & 0.98 & 0 & $-0.2$ & $^{+1.7}_{-1.5}$ & 0 & $-0.3$ & $^{+0.9}_{-0.8}$ & $0.5^{+0.8}_{-1.2}$ \\
\hline
 & $B-V$ & 0.99 & 0 & $-2.3$ & $^{+1.4}_{-1.4}$ & 0 & $-0.3$ & $^{+0.4}_{-0.4}$ & $2.2^{+0.7}_{-2.7}$ \\
59111 & $B-R$ & 0.99 & 0 & $0.1$ & $^{+1.4}_{-1.3}$ & 0 & $0.0$ & $^{+0.4}_{-0.4}$ & $1.1^{+0.7}_{-1.9}$ \\
 & $V-R$ & 0.99 & 0 & $2.4$ & $^{+1.4}_{-1.4}$ & 0 & $0.7$ & $^{+0.3}_{-0.3}$ & $0.5^{+0.7}_{-1.0}$ \\
\hline
 & $B-V$ & 0.83 & 0 & $0.3$ & $^{+1.0}_{-1.0}$ & 0 & $0.4$ & $^{+0.7}_{-0.7}$ & $0.3^{+1.6}_{-1.6}$ \\
59887 & $B-R$ & 0.79 & 0.3 & $-1.0$ & $^{+0.9}_{-1.1}$ & 0 & $-1.8$ & $^{+0.6}_{-0.6}$ & $0.5^{+1.5}_{-3.1}$ \\
 & $V-R$ & 0.83 & 0 & $-1.9$ & $^{+0.9}_{-0.8}$ & 0 & $-2.3$ & $^{+0.5}_{-0.6}$ & $0.2^{+1.0}_{-1.3}$ \\
\hline
 & $B-V$ & 0.94 & 0 & $-4.1$ & $^{+1.7}_{-1.7}$ & 0 & $-1.4$ & $^{+0.4}_{-0.4}$ & $0.3^{+2.6}_{-3.6}$ \\
59888 & $B-R$ & 0.94 & 0 & $-4.0$ & $^{+1.8}_{-1.7}$ & 0 & $-1.5$ & $^{+0.4}_{-0.4}$ & $0.5^{+1.6}_{-3.0}$ \\
 & $V-R$ & 0.97 & 0 & $-0.4$ & $^{+1.7}_{-1.7}$ & 0 & $-0.1$ & $^{+0.3}_{-0.3}$ & $0.2^{+0.6}_{-1.4}$ \\
\enddata
\tablecomments{The columns stand for observational date (MJD), bands, the maximum correlation coefficient in ICCF, removal rate in ICCF and \texttt{PyROA}, the measured lag of ICCF, \texttt{PyROA} and ZDCF, respectively. Positive values indicate that the former band leads the latter one.}
\end{deluxetable*}

We performed the cross-correlation analysis to measure possible interband time lags, using three types of time series analysis approaches: the interpolated cross-correlation function \citep[ICCF,][]{1998ApJ...501...82P, 1998PASP..110..660P}, \texttt{PyROA}\footnote{\url{https://github.com/FergusDonnan/PyROA}}\citep{2021MNRAS.508.5449D}, and the $z$-transformed discrete correlation function \citep[ZDCF,][]{1997ASSL..218..163A, 2013arXiv1302.1508A}. The search range of the time lags was set as $\pm60$ minutes. Combining linear interpolation and cross-correlation analysis, ICCF estimates the optimal time lag by calculating the Pearson coefficient $r$ between the two light curves at each given time shift $\tau$. It was implemented in this paper through the publicly available algorithm \texttt{PyCCF}\footnote{\url{http://ascl.net/code/v/1868}}\citep{1998PASP..110..660P,2018ascl.soft05032S}. Based on the running optimal average (ROA), \texttt{PyROA} models AGN light curves and measures time lags, and parameters are sampled using Markov Chain Monte Carlo (MCMC) techniques. In addition, priors can be used on the sampled parameters by means of a Bayesian approach. 

\begin{figure*}[ht!]
\plotone{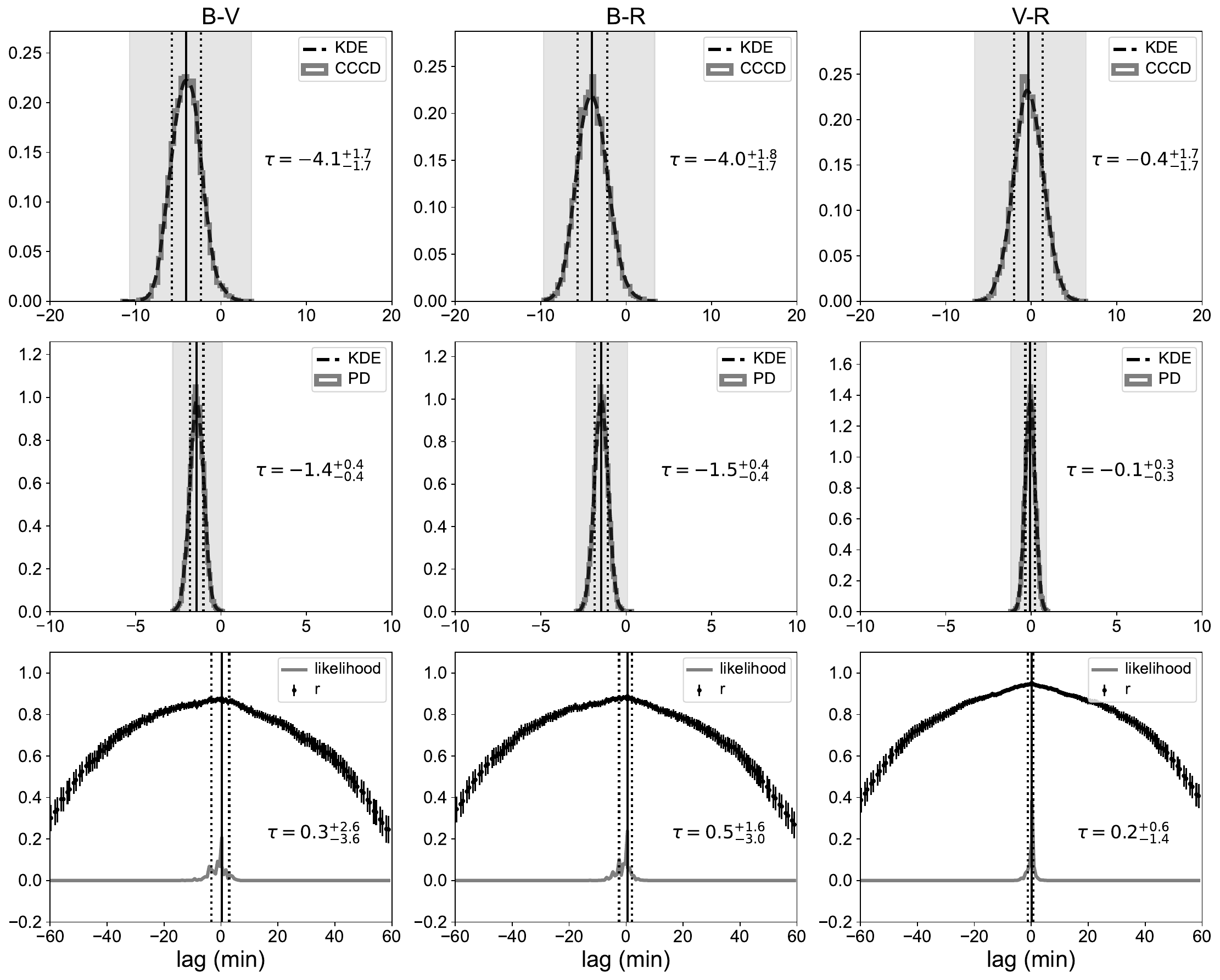}
\caption{Examples of lags measured by three approaches on MJD 59888. The left, center, and right columns represent lags in the $B-V$, $B-R$, and $V-R$ bands, respectively. Lags and uncertainties are given in all panels and are labeled by the solid and dotted lines, respectively. Top two rows: the CCCD/PD obtained from ICCF/\texttt{PyROA}. The dashed lines are the KDE of the weighted CCCD/PD and the histograms are the unweighted ones. The shaded regions are the primary peaks. The x-axes have been zoomed in for clarity. Bottom row: the discrete correlation coefficient (points) and the likelihood (curves) obtained from ZDCF and \texttt{PLIKE}, respectively.
\label{fig5}}
\end{figure*}

To estimate the final time lag and its uncertainties, 10,000 Monte Carlo (MC) realizations were applied to derive the cross-correlation centroid distribution (CCCD) for ICCF, while 10,000 MCMC iterations were conducted for \texttt{PyROA} to obtain the posterior distribution (PD) of the time lag. ICCF-derived time lags served as the prior for the initialization of $\tau$ in \texttt{PyROA}. To address the issue of multiple peak alias in the CCCD/PD, also mentioned in previous articles \citep[e.g.,][]{2017ApJ...851...21G, 2019ApJ...884..119L, 2024ApJS..272...26S}, we subsequently adopted the alias-removal procedure described in \citet{2017ApJ...851...21G}. The procedure smoothed the weighted lag distributions in the CCCD/PD through a Gaussian kernel density estimate (KDE). The weights of the distributions were expressed as $P = [N(\tau)/N(0)]^2$, where $N(\tau)$ represents the number of data points that overlap between the two light curves with a time shift $\tau$. Only the peak containing the largest area in the distributions were kept for lag measurement. The median and 16th/84th percentiles of the initial distributions within the primary peak were identified as the time lag and lower/upper limits, respectively.

The ZDCF analysis was carried out making use of \texttt{pyPETaL}\footnote{\url{https://ascl.net/2401.004}}\citep{2024ascl.soft01004S}, which provides a Python implementation and wrapper for ZDCF and was originally introduced in \citet{2024ApJS..272...26S} for lag estimation. In ZDCF, the data were grouped into equal population bins, followed by the application of Fisher $z$-transformation to reduce the skewness in the distribution of the correlation coefficient. The ZDCF was derived by performing 10,000 MC realizations. The final time lag and the associated uncertainties were estimated using the \texttt{PLIKE} algorithm \citep{2013arXiv1302.1508A}, which determines the maximum likelihood value and the 68.2\% confidence interval of the normalized likelihood distribution based on the likelihood function.

The complete results are displayed in Table~\ref{tab2}, and the examples of time lags on MJD 59888 obtained by the three approaches are presented in Figure~\ref{fig5}. Positive lags indicate that the variation in the former band leads that in the latter one. It is worth emphasizing that the Gaussian smoothing applied in \texttt{PyROA} may introduce overfitting, potentially leading to underestimated confidence intervals. For this case, we could manually adjust the initial range of $\Delta$ to mitigate it \citep{2024ApJS..275...13W}. In contrast, the results given by ZDCF sometimes have relatively larger uncertainties. Among the three methods, ICCF provides the most robust measurements \citep{2024ApJS..275...13W}. Therefore, for the analysis of time lags, we primarily adopt ICCF as our main estimator, while also taking into account the outputs from the other two methods for cross-validation.

From Table~\ref{tab2}, it can be seen that most of the measured time lags are close to zero, although the non-zero results given by ICCF in the $B-V$ and $V-R$ bands on MJD 59111, as well as in the $B-V$ and $B-R$ bands on MJD 59888, correspond to the spectral hysteresis described in Section~\ref{3.3}. We reject these non-zero time lags measured by ICCF due to the inconsistency between the results derived from three methods. The absence of time lags between variations in optical bands across diverse timescales is common \citep[e.g.,][]{2018ApJ...862..123M, 2022ApJ...933..224F, 2023ApJ...943..135K, 2025MNRAS.537.2586A}. Up to now, there have been only a few reports of time lag detection in the IDV of BL Lac. For example, \citet{2022ApJ...926...91F} detected time lags of approximately 16 minutes in the $B-V$ bands and 18 minutes in the $B-R$ bands, accompanied by two clockwise loops. As discussed in \citet{2012AJ....143..108W}, the wavelength separations between the optical bands are too small and make lag detection challenging, suggesting that the optical emission within the jet is cospatial. Moreover, given that the optical interband time lags may occur on minute timescales, a temporal resolution of $\sim 1$ min is ideal for the detection of these time lags. In the future, a higher temporal resolution and an increase in the number of exposures will facilitate the exploration of time lags among the optical bands \citep{2017MNRAS.469.3588M, 2025MNRAS.537.2586A}.

\section{Long-term Variability Analysis} \label{4}

\subsection{Flux Variation Study} \label{4.1}

\begin{figure*}[ht!]
\plotone{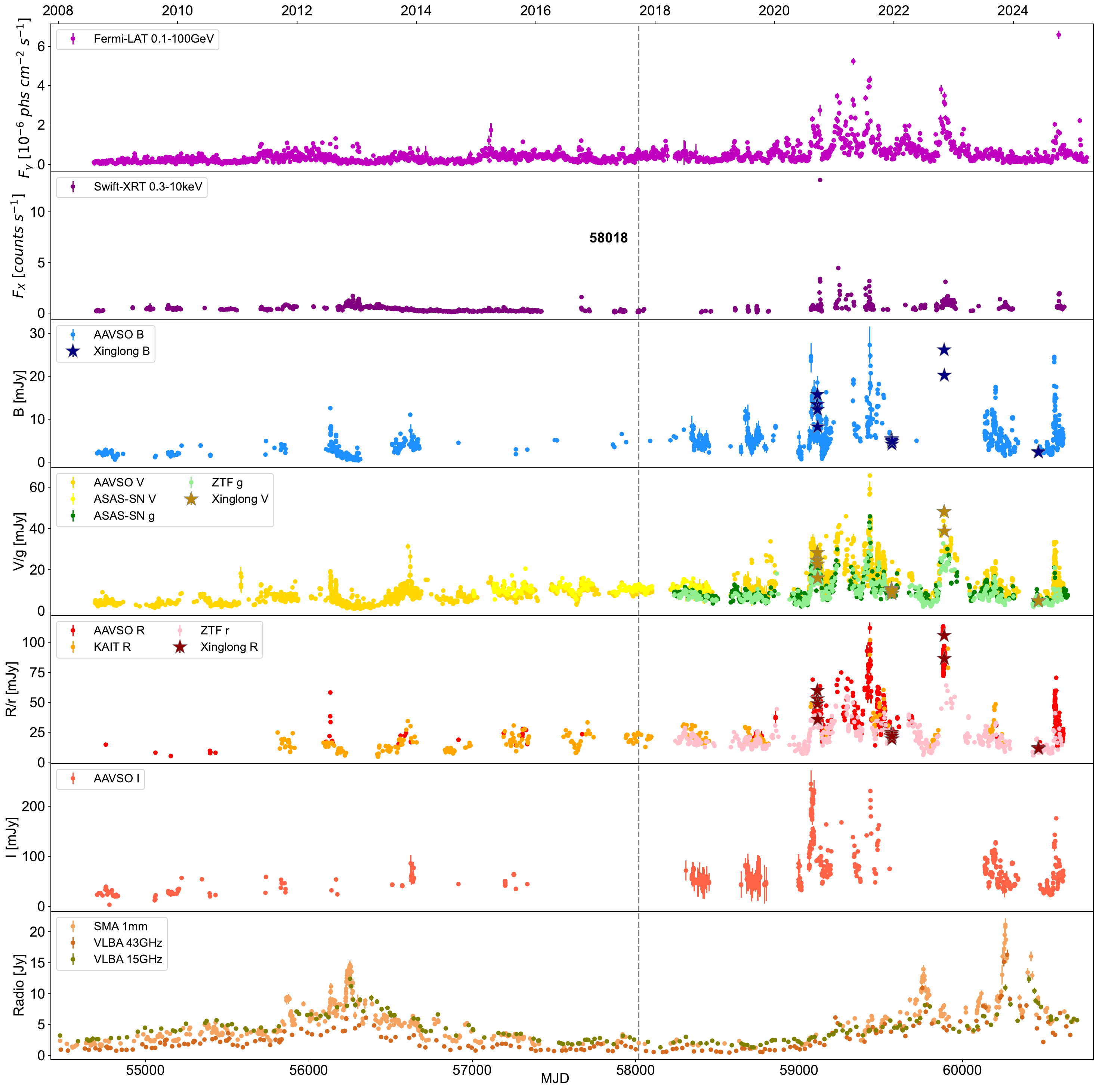}
\caption{The multiwavelenth light curves of BL Lac covering over 17 years. Top two panels: the light curves of 0.1-100 GeV ($\gamma$-ray) and 0.3--10 keV (X-ray). From third to sixth panels: the multiband optical light curves gathered from multiple monitoring missions. Asterisks in these panels represent the nightly-mean magnitudes derived from our 12 night observations. The bottom panel plots the radio light curves of 1 mm, 43 GHz and 15 GHz. The dashed line indicates the date selected for SED fitting in Section~\ref{5}, with MJD labeled alongside the line.
\label{fig6}}
\end{figure*}

Figure~\ref{fig6} displays the long-term multiwavelength light curves of BL Lac observed by all the facilities introduced in Section~\ref{2}. The star-shaped markers are the nightly-mean magnitudes computed from the 12 night observations at Xinglong Station and indicate a strong flare in the $B$-band that was missed in the AAVSO data.

\begin{figure}[ht!]
\begin{center}
    \includegraphics[angle=0,scale=0.32]{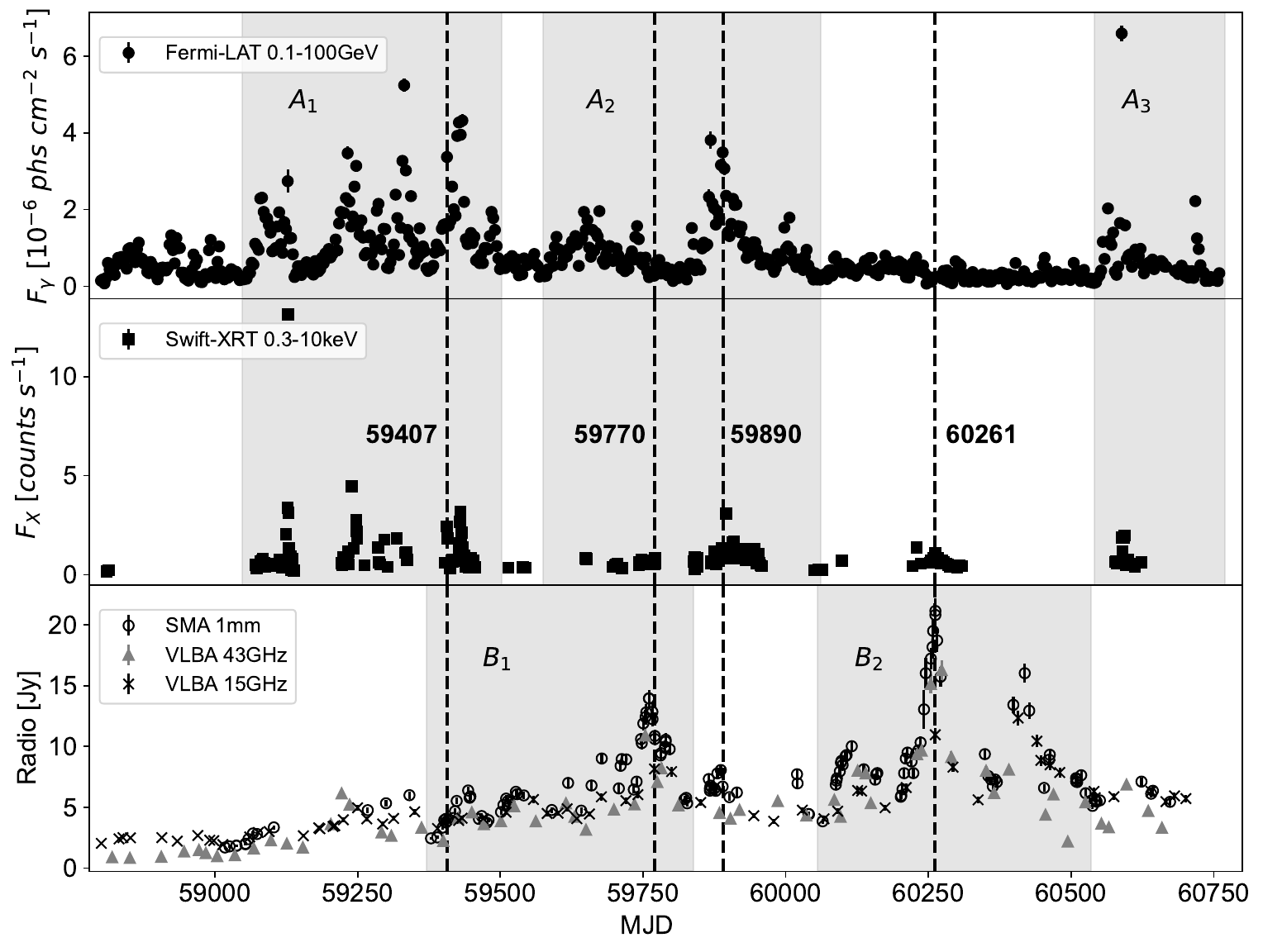}
    \caption{The $\gamma$-ray, X-ray and radio light curves of BL Lac after MJD 58780. The shaded regions represent high-energy flaring episodes $\rm A_1$, $\rm A_2$, and $\rm A_3$, and radio flaring episodes $\rm B_1$ and $\rm B_2$, respectively. The dashed lines indicate the dates selected for SED fitting in Section~\ref{5}, with MJDs labeled alongside the lines.
    \label{fig7}}
\end{center}
\end{figure}

The multiwavelength light curves demonstrate the complex nature of LTV in BL Lac. Initially, the source remained in a quiescent state until around MJD 55700 (i.e., approximately 2011 May), when it entered a flaring phase. During this flare, the flux enhancement in the optical and radio bands was more pronounced than that in the $\gamma$-ray and X-ray energies. Subsequently, the source returned to a quiescent state that lasted for about seven years. During this quiescent period, a few minor flares were detected in the $\gamma$-ray and optical bands, whereas almost no flaring activity was observed in the X-ray and radio bands. Beginning on about MJD 58850 (i.e., roughly 2020 January), BL Lac underwent an exceptionally active phase, clearly manifested in all bands. This active phase even surpassed that observed after MJD 55700, especially in the $\gamma$-ray and X-ray energies. Furthermore, it is readily observed that the flares in the radio bands occurred later than those in the optical to $\gamma$-ray energies. In comparison, for the flares observed after MJD 55700, visual inspection suggests that the variations in all bands do not show obvious time lag. To better investigate the flux variations of BL Lac during this period, we segmented the high-energy and radio light curves into several flaring episodes. The $\gamma$-ray and X-ray flaring episodes are denoted as episodes $\rm A_1$ (from MJD 59040 to 59500), $\rm A_2$ (from MJD 59570 to 60060), and $\rm A_3$ (from MJD 60540 to 60760), while radio flaring episodes are denoted as episodes $\rm B_1$ (from MJD 59370 to 59840) and $\rm B_2$ (from MJD 60050 to 60540), highlighted by the shaded regions in Figure~\ref{fig7}. 

During episodes $\rm A_1$, $\rm A_2$, and $\rm A_3$, the maximum $\gamma$-ray fluxes observed in the light curve were $(5.24\pm0.17)\times10^{-6}\ \mathrm{phs\ cm^{-2}\ s^{-1}}$ on MJD 59331.5, $(3.81\pm0.22)\times10^{-6}\ \mathrm{phs\ cm^{-2}\ s^{-1}}$ on MJD 59868.5, and $(6.59\pm0.21)\times10^{-6}\ \mathrm{phs\ cm^{-2}\ s^{-1}}$ on MJD 60588.5, respectively. Due to the differences in the selected energy range and the temporal binning method, the flux values on MJD 59331.5 and MJD 59868.5 differ from those reported in \citet{2024MNRAS.527.5140S}, but both represent the peak fluxes detected in episodes $\rm A_1$ and $\rm A_2$. In particular, the flux measured on MJD 60588.5 marks the highest $\gamma$-ray emission from BL Lac ever recorded by Fermi-LAT. According to \citet{2024ATel16849....1V}, the daily averaged $\gamma$-ray flux ($E>100\ \rm MeV$) of BL Lac reached $(10.4\pm0.5)\times10^{-6}\ \mathrm{phs\ cm^{-2}\ s^{-1}}$ on that date.  
In addition, this VHE emission from BL Lac was quasi-simultaneously captured by several VHE telescopes, such as MAGIC, VERITAS and LHAASO \citep{2024ATel16861....1P, 2024ATel16854....1C, 2024ATel16850....1X}. Similarly, the X-ray emission of BL Lac shows multiple peak fluxes within the three episodes. The brightest state of the source was observed on MJD 59128.18, with a count rate of $13.12\ \rm counts\  s^{-1}$, which matches the findings of \citet{2021MNRAS.507.5602P}. Studies on the SED of BL Lac on that day propose that its synchrotron peak frequency $\nu_{\rm p}^{\rm sync}$ rose above $10^{15}\ \mathrm{Hz}$, implying that the source temporarily behaved like an HBL \citep{2021MNRAS.507.5602P, 2025MNRAS.536.1251W}.

\begin{deluxetable}{lcccccc}[t!]
\tablewidth{0pt}
\tabletypesize{\footnotesize}
\tablecaption{Mean and Peak Fluxes in Radio Flaring Episodes\label{tab3}}
\tablehead{
\colhead{} & \multicolumn{3}{c}{Episode $\rm B_1$} & \multicolumn{3}{c}{Episode $\rm B_2$} \\
\cmidrule(lr){2-4} \cmidrule(lr){5-7}
\colhead{Band} & \colhead{$F_{\rm mean}$} & \colhead{Date} & \colhead{$F_{\rm peak}$} & \colhead{$F_{\rm mean}$} & \colhead{Date} & \colhead{$F_{\rm peak}$} \\
\colhead{} & \colhead{(Jy)} & \colhead{(MJD)} & \colhead{(Jy)} & \colhead{(Jy)} & \colhead{(MJD)} & \colhead{(Jy)} 
}
\startdata
1 mm & 7.173 & 59760 & 13.951 & 9.736 & 60262 & 21.139 \\
43 GHz & 5.091 & 59754 & 10.896 & 7.696 & 60273 & 16.278 \\
15 GHz & 5.095 & 59770 & 8.153 & 7.480 & 60407 & 12.336 \\
\enddata
\tablecomments{The columns are radio band, mean flux, peaked date (MJD) and peak flux in each episode, respectively.}
\end{deluxetable}

\begin{deluxetable*}{l c D{.}{.}{2.1} r@{}l D{.}{.}{2.1} r@{}l r@{}l}[ht!]
\tablewidth{0pt}
\tabletypesize{\normalsize}
\tablecaption{Results of Broad-band Time Lags\label{tab4}}
\tablehead{
\colhead{Bands} & \multicolumn{4}{c}{ICCF} & \multicolumn{3}{c}{\texttt{PyROA}} & \multicolumn{2}{c}{ZDCF} \\
\cmidrule(lr){2-5} \cmidrule(lr){6-8} \cmidrule(lr){9-10}
\colhead{} & \colhead{$r_{\rm max}$} & \multicolumn{1}{c}{$f_{\rm out}$ (\%)} & \multicolumn{2}{c}{Lag (d)} & \multicolumn{1}{c}{$f_{\rm out}$ (\%)} & \multicolumn{2}{c}{Lag (d)} & \multicolumn{2}{c}{Lag (d)}
}
\startdata
$\gamma$-ray vs. X-ray & 0.53 & 1.0 & $2.4$ & $^{+5.4}_{-2.4}$ & 4.5 & $2.3$ & $^{+0.6}_{-0.8}$ & $72.2$ & $^{+45.1}_{-124.8}$ \\
$\gamma$-ray vs. $g$ & 0.83 & 0.2 & $-0.4$ & $^{+1.4}_{-0.8}$ & 4.5 & $0.9$ & $^{+0.2}_{-0.1}$ & $-2.1$ & $^{+15.0}_{-13.5}$ \\
$\gamma$-ray vs. 1 mm & 0.55 & 0.4 & $364.2$ & $^{+5.8}_{-5.9}$ & 13.6 & $374.1$ & $^{+25.9}_{-43.3}$ & $375.0$ & $^{+23.0}_{-31.8}$ \\
$\gamma$-ray vs. 43 GHz & 0.60 & 0.7 & $366.7$ & $^{+9.3}_{-9.6}$ & 4.4 & $389.7$ & $^{+55.3}_{-44.9}$ & $358.1$ & $^{+16.9}_{-8.6}$ \\
$\gamma$-ray vs. 15 GHz & 0.45 & 10.3 & $375.3$ & $^{+13.7}_{-6.6}$ & 31.8 & $365.3$ & $^{+40.8}_{-30.7}$ & $437.8$ & $^{+32.3}_{-59.9}$ \\
X-ray vs. $g$ & 0.37 & 25.4 & $-2.0$ & $^{+3.0}_{-13.3}$ & 0.0 & $1.3$ & $^{+1.2}_{-1.4}$ & $-4.0$ & $^{+24.2}_{-7.6}$ \\
X-ray vs. 1 mm & 0.45 & 4.0 & $362.6$ & $^{+3.1}_{-7.6}$ & 45.1 & $334.3$ & $^{+3.4}_{-3.2}$ & $435.7$ & $^{+24.6}_{-129.9}$ \\
X-ray vs. 43 GHz & 0.52 & 11.9 & $357.8$ & $^{+4.7}_{-10.2}$ & 4.4 & $356.8$ & $^{+1.0}_{-0.8}$ & $355.7$ & $^{+31.7}_{-28.6}$ \\
X-ray vs. 15 GHz & 0.45 & 42.8 & $364.2$ & $^{+7.3}_{-84.6}$ & 0.0 & $362.1$ & $^{+3.6}_{-35.6}$ & $383.5$ & $^{+71.4}_{-29.0}$ \\
$g$ vs. 1 mm & 0.55 & 0.1 & $362.6$ & $^{+4.4}_{-5.9}$ & 4.6 & $368.5$ & $^{+37.5}_{-79.4}$ & $373.2$ & $^{+10.8}_{-25.0}$ \\
$g$ vs. 43 GHz & 0.58 & 0.5 & $349.0$ & $^{+9.0}_{-7.5}$ & 0.0 & $353.8$ & $^{+3.5}_{-3.5}$ & $371.2$ & $^{+18.2}_{-30.5}$ \\
$g$ vs. 15 GHz & 0.40 & 2.2 & $370.2$ & $^{+19.2}_{-10.4}$ & 9.1 & $360.7$ & $^{+62.0}_{-65.0}$ & $351.1$ & $^{+63.1}_{-84.7}$ \\
1 mm vs. 43 GHz & 0.94 & 0.0 & $-15.8$ & $^{+9.2}_{-9.4}$ & 4.1 & $12.8$ & $^{+0.1}_{-0.2}$ & $12.4$ & $^{+17.4}_{-29.8}$ \\
1 mm vs. 15 GHz & 0.91 & 48.3 & $4.9$ & $^{+5.5}_{-5.3}$ & 0.0 & $17.2$ & $^{+2.2}_{-1.5}$ & $15.2$ & $^{+19.4}_{-18.3}$ \\
43 GHz vs. 15 GHz & 0.93 & 0.0 & $39.9$ & $^{+34.3}_{-37.0}$ & 13.5 & $27.4$ & $^{+0.2}_{-0.3}$ & $-4.8$ & $^{+27.2}_{-23.7}$ \\
\enddata
\tablecomments{The columns stand for bands, the maximum correlation coefficient in ICCF, removal rate in ICCF and \texttt{PyROA}, the measured lag of ICCF, \texttt{PyROA} and ZDCF, respectively. Positive values imply that the former band leads the latter one.}
\end{deluxetable*}

Throughout episodes $\rm B_1$ and $\rm B_2$, radio fluxes exhibited significant enhancements and reached clear peaks at all three frequencies. Their respective mean fluxes and peak fluxes are summarized in Table~\ref{tab3}. During episode $\rm B_2$, the light curves of the three radio bands show three distinct peaks. The 1 mm light curve reached its peak flux on MJD 60262, and the peak flux of 43 GHz appeared on MJD 60273. However, the peak flux of 15 GHz occurred on MJD 60407. The profiles of the light curves at the three frequencies are not completely consistent, which could be attributed to the sparse observational sampling that may have missed some detailed variations. The peak fluxes show a positive correlation with the observing frequencies, reflecting that the variability amplitudes become larger at higher frequencies, whereas the light curves appear to be more smoothed toward longer wavelengths. This trend was frequently observed previously. As an example, \citet{2015A&A...582A.103G} reported a trend of an increase in variation amplitude with frequency in the variability of BL Lac from 4.8 GHz to 36.8 GHz. For the light curves after MJD 58850, when excluding the data points of episodes $\rm B_1$ and $\rm B_2$, the average flux of the remaining data points stayed above the low flux level observed before MJD 58850. This suggests that the post-MJD 58850 light curve consists of a long-term trend with an elevated baseline flux, upon which year-scale flaring events are superimposed. This kind of overlapping feature was also present in the radio outburst after MJD 55700 \citep{2025MNRAS.536.1251W}. The elevated baseline flux can be attributed to the relatively flat and slow variability in the radio domain, which evolves over longer timescales. \citet{2024A&A...692A..48R} computed the auto-correlation function of the 15 GHz light curve of BL Lac and derived a characteristic timescale of 95--120 days, based on the data spanning approximately MJD 58450--59650. The source may have continuously produced a series of minor flares with this characteristic timescale that overlapped with each other, eventually leading to the elevated baseline flux observed after MJD 58850.

\subsection{Broad-band Time Lags and Emission Regions} \label{4.2}

\begin{figure*}[ht!]
\plotone{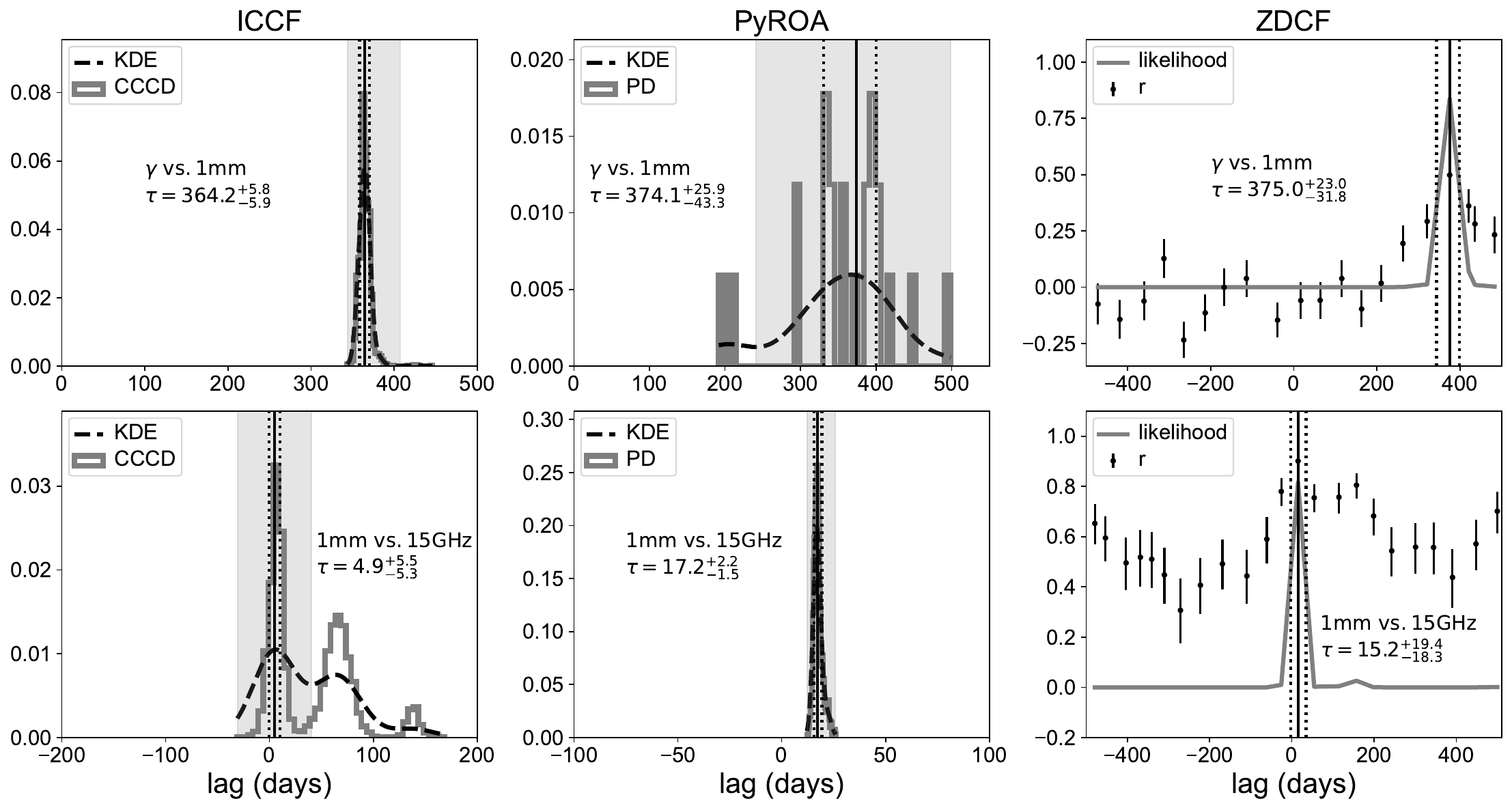}
\caption{Two examples of broad-band time lags measured by three approaches. The top and bottom rows represent lags between $\gamma$-ray and 1 mm, and between 1 mm and 15 GHz, respectively. The left, center, and right columns are the CCCD, PD, and discrete correlation coefficient/likelihood curve obtained from ICCF, \texttt{PyROA}, and ZDCF/\texttt{PLIKE}, respectively. The x-axes in the left and center columns have been zoomed in for clarity.
\label{fig8}}
\end{figure*}

As described in the previous subsection, the flares in the radio bands of BL Lac exhibit delays relative to those in other wavelengths, suggesting different locations of their emission regions in the jet. This can be clarified with correlation analysis. We employed the same cross-correlation analysis procedure as in Section~\ref{3.4} and derived the time lags between the light curves of $\gamma$-ray, X-ray, optical $g$-band, 1 mm, 43 GHz and 15 GHz, with a search range of $\pm500$ days. The optical $g$-band light curve was from ZTF observations, which started from MJD 58234.5. Therefore, all light curves used in the cross-correlation analysis were truncated as from MJD 58230 for consistency. All time lag measurements are listed in Table~\ref{tab4}, and two representative cases of the cross-correlation analysis are shown in Figure~\ref{fig8}. We still adopt ICCF as our main estimator and the other two methods for cross-validation as mentioned in Section~\ref{3.4}.

The variations of three pairs of bands exhibit time lags close to zero: $\gamma$-ray versus X-ray, $\gamma$-ray versus $g$-band, and X-ray versus $g$-band. The time lag between $\gamma$-ray and X-ray derived by ZDCF deviates from the results obtained by the other two methods. \citet{2024MNRAS.527.5140S} measured the time lag using ZDCF and acquired a value of $-92.16^{+90.12}_{-3.60}$ days based on the data from MJD 59000 to 59943. This situation is likely attributable to the sparse and irregular sampling of the X-ray data. Taking into account the values and uncertainties of all three methods, the time lag between $\gamma$-ray and X-ray is considered to be approximately zero. In addition, \citet{2022Natur.609..265J} and \citet{2024A&A...692A..48R} found the strong correlation between $\gamma$-ray and $R$-band at $\tau \sim 0$. The lags between 1 mm, 43 GHz and 15 GHz bands mean that the variations at different radio frequencies have inter-band time lags of days to weeks, and most of our results suggest that the variations at higher frequencies lead those at lower frequencies. The lags between the radio bands have likewise been mentioned in other works \citep[e.g.,][]{2011MNRAS.415.1631K, 2015A&A...582A.103G}. According to \citet{2015A&A...582A.103G}, for the light curves from MJD 55200 to 56500, the variation of BL Lac at 36.8 GHz occurred earlier than that at lower frequencies, with lead times ranging from about 2.11 to 25.76 days. Nevertheless, the time lags between different radio frequencies are deemed minor and thus negligible compared to the considerable lags detected between radio and those bands at higher frequencies in the following discussion. On the other hand, significant inter-band time lags were detected between the radio bands and the optical to $\gamma$-ray energies. The ICCF analyses indicate time lags ranging from 349.0 to 375.3 days. An average value of about 370 days is adopted for further computations.

Our lag measurements for the outburst after MJD 58850 exhibit some differences from that after MJD 55700, which deserve discussion. For the outburst after MJD 55700, \citet{2013MNRAS.436.1530R} reported that the millimeter emission lagged behind the $\gamma$-ray/$R$-band emission by 120--150 days, while the correlation between the X-ray and millimeter flux variations was probably consistent with zero lag. \citet{2025MNRAS.536.1251W} determined a time lag of $2.35^{+8.25}_{-6.80}$ days between the X-ray and 15 GHz, and $-167.9^{+6.05}_{-7.70}$ days between the X-ray and $V$-band. In addition, \citet{2015A&A...582A.103G} measured the time lag between the R-band and 36.8 GHz light curves, obtaining a value of $250.28\pm10.21$ days. These results suggest that during the outburst after MJD 55700, the X-ray and radio emissions arise from the same region, while the $\gamma$-ray and optical emissions from another shared region. However, the X-ray light curve monitored by Swift lacks extensive sampling in 2012, while the Rossi X-ray Timing Explorer (RXTE) Proportional Counter Array (PCA) ceased observations on MJD 55926 \citep[i.e., 2011 December 31,][]{2013MNRAS.436.1530R, 2016ApJ...816...53W}, making it difficult to determine the specific variability features of BL Lac in the X-ray band during 2012. For the flaring event after MJD 58850, our calculations indicate that the emissions from optical to $\gamma$-ray wavelength originate from the same radiation zone, whereas the 1 mm to 15 GHz emissions can be attributed to a more extended radio emission zone. The time lags between the two emission zones are also found to be larger than those observed after MJD 55700.

Building upon the scenario of two emission zones in the jet, a shock propagating along the jet can further explain the measured time lags. \citet{2011MNRAS.415.1631K} provides the formula for estimating the distance between distinct emission regions: 
\begin{equation} \label{eq3}
\Delta d = \frac{\beta_{\rm app}c\Delta t}{(1+z)\rm sin\theta},
\end{equation}
where $\beta_{\rm app}$ is the apparent speed in units of $c$, $c$ is the speed of light, $\Delta t$ is the measured time lag, $z$ is the redshift, $\theta$ is the viewing angle between the jet orientation and line of sight. Substituting the average $\beta_{\rm app}$ of 4.46 reported by \citet{2021ApJ...923...30L}, an average time lag of 370 days, the redshift of BL Lac ($z=0.0688$), and the viewing angle of $5.1^{\circ}$
from \citet{2017ApJ...846...98J} into the equation above yields a distance $\Delta d$ of roughly 14.58 pc, i.e., $4.50\times10^{19}\ \rm cm$. According to the model constructed in \citet{2008Natur.452..966M}, the radio emission region is expected to lie beyond $10^5$ times the Schwarzschild radius ($R_{\rm S}$). \citet{2018ApJS..235...39C} provides the central supermassive black hole mass for BL Lac of $10^{8.21}\ \rm M_{\astrosun}$, yielding a $R_{\rm S}$ of $4.79\times10^{13}\ \rm cm$. Based on this, our calculated radio emission region is located at a distance of approximately $9.39\times10^{5}\ R_{\rm S}$ from the central engine.

\section{Broad-band SED Modeling} \label{5}

\subsection{Model Construction and Parameter Settings} \label{5.1}

Extensive studies have been devoted to the SED fitting of BL Lac. Almost all of them interpret the double-humped structure of BL Lac in the framework of a leptonic model. However, a single-zone leptonic model sometimes fails to fully reproduce the SED, making a two-zone leptonic model necessary. In some studies, the IC process of BL Lac is dominated by the SSC radiation \citep{2011ApJ...730..101A, 2016ApJ...816...53W}. Some spectroscopic studies have reported the detection of weak emission lines, e.g., $\rm H\alpha$ and $\rm H\beta$ lines, suggesting the presence of a BLR in this source \citep{2010A&A...516A..59C, 2021ApJS..253...46P}. Consequently, the BLR could also contribute seed photons for the IC process. A scenario in which the IC emission comprises both the SSC and the EC components is reasonable. 
Although the unified model of AGN theoretically predicts the existence of a dusty torus (DT) in blazars, direct observational signatures for such a structure in BL Lac objects is lacking \citep{2012ApJ...745L..27P}. The EC scattering on seed photons from the DT is commonly considered in the SED modeling of FSRQs, such as 3C 454.3 \citep{2008ApJ...675...71S} and PKS 1510--089 \citep{2017A&A...603A..29A}. Therefore, we excluded the EC contribution of DT photons from our model in this work.
Numerous papers have successfully modeled the SED of BL Lac using the SSC+EC scenario \citep[e.g.,][]{2011ApJ...730..101A, 2019A&A...623A.175M, 2022MNRAS.513.4645S, 2024MNRAS.527.5140S}, where the EC component arises from the scattering of soft photons from the BLR. As discussed at the end of Section~\ref{4.2}, the radio emission region is located downstream along the jet than the high-energy region. This supports a preliminary assumption that the IC process in the high-energy region involves both SSC and EC mechanisms, whereas the IC process in the radio emission region is primarily driven by SSC alone.

To investigate the temporal evolution of the SED during the multiwavelength flaring stage of BL Lac, we selected the data on those dates with high-energy and/or radio flares for SED modeling. To ensure that those data from each band could effectively constrain the SED modeling, multiwavelength observations with time lags in the range of 349.0--375.3 days were selected, and quasi-simultaneous data were used in the fitting. As a result, four epochs were chosen: MJDs 59407, 59770, 59890, and 60261, which correspond to the flaring peaks in episodes $\rm A_1$, $\rm A_2$, $\rm B_1$, and $\rm B_2$, respectively (see Figure~\ref{fig7}). 
Specifically, the radio flare on MJD 59770 is identified as the counterpart to the high-energy flare on MJD 59407, and a similar association applies to the latter pair. We propose that each pair of flares corresponds to the genesis and downstream propagation of a new shock within the jet. According to the internal shock model \citep{2001MNRAS.325.1559S, 2010ApJ...711..445B}, when a faster plasma shell ejected by the central engine overtakes a slower preceding one, the collision generates a shock. This shock accelerates the plasma and enhances the physical conditions of the emission region, triggering a flare. Consequently, the shock first dissipates energy in the inner zone, producing the high-energy flare, and subsequently propagates along the jet to the radio zone to induce the radio flare. This physical scenario naturally explains the time lag of about one year between the high-energy and radio flares.

In addition, MJD 58018, when BL Lac was in a quiescent state, was also chosen for the comparative analysis (see Figure~\ref{fig6}). 
For the quiescent state (MJD 58018) and the high-energy flaring states (MJDs 59407 and 59890), we assumed that the radiation from BL Lac was dominated by the high-energy region. Conversely, for the radio flaring states (MJDs 59770 and 60261), the SEDs were modeled as being dominated by the radio emission region. Although VLBA observations indicate that the radio zone is a persistent feature, this study focuses on the physical properties within this region during radio flares. Therefore, to reduce model complexity, we did not explicitly include the emission from the radio zone in the fitting for MJDs 58018, 59407, and 59890. This simplification does not imply the physical disappearance of this region at those times.
The magnitudes of the selected optical data were corrected for Galactic extinction using the values provided by the NASA/IPAC Extragalactic Database\footnote{\url{https://ned.ipac.caltech.edu/}}(NED), and then converted into flux densities following the procedure described in \citet{1998A&A...333..231B}.

In this work, the broad-band SED of BL Lac was modeled under the leptonic scenario using the publicly available code \texttt{JetSeT} version 1.3.0\footnote{\url{https://jetset.readthedocs.io/en/latest/}}\citep{2009A&A...501..879T, 2011ApJ...739...66T, 2020ascl.soft09001T}. This code carries out the SED fitting and computes the best-fit parameters by means of the Minuit optimizer and MCMC sampling. The physical parameters of the emission region are based on the following assumptions: the emission region is approximated as a spherical blob with a radius $R$ located at a distance $D$ from the central engine, moving with a bulk Lorentz factor $\Gamma$ at a small angle $\theta$ with respect to the line of sight. It contains a magnetic field of intensity $B$, and a population of relativistic electrons whose energy distribution follows a power law with an exponential cutoff given by
\begin{equation} \label{eq4}
N(\gamma) = N_e\gamma^{-p}\rm exp(-\frac{\gamma}{\gamma_{\rm cut}}),
\end{equation}
where $\gamma_{\rm min}\leq\gamma\leq\gamma_{\rm max}$, in which $\gamma_{\rm min}$, $\gamma_{\rm max}$, and $\gamma_{\rm cut}$ are the minimum, maximum, and cutoff energies of the electrons, respectively, and $p$ is the spectral index of the power law. 
The bulk Lorentz factor $\Gamma$ and the viewing angle $\theta$ together determine the Doppler factor $\delta$ with the definition of $\delta=1/\Gamma(1-\beta\rm cos\theta)$, where $\beta$ is the speed of the blob in units of $c$. $\Gamma$ is a function of $\beta$, given by $\Gamma=(1-\beta^2)^{-1/2}$. A larger $\Gamma$ and a smaller $\theta$ result in a higher $\delta$, leading to a stronger Doppler boosting effect.
Additional parameters involved in the SED modeling, including the accretion disk luminosity $L_{\rm disk}=3.31\times10^{43}\ \rm erg\ s^{-1}$ and the disk temperature $T_{\rm disk}=1.16\times10^5\ \rm K$, are determined based on the values from the literatures \citep{2018ApJS..235...39C, 2025ApJ...978...43M}. The BLR is treated as a spherical shell with finite thickness. Its inner and outer radii depend on $L_{\rm disk}$, and are defined by the following formulae given by \texttt{JetSeT}:
\begin{equation} \label{eq5}
R_{\rm BLR,in} = 3 \times 10^{17}(\frac{L_{\rm disk}}{10^{46}})^{0.5}=1.72\times10^{16}\ \rm cm,
\end{equation}
\begin{equation} \label{eq6}
R_{\rm BLR,out} = 1.1 \times R_{\rm BLR,in}=1.90\times10^{16}\ \rm cm.
\end{equation}
The parameters associated with the accretion disk and the BLR are kept fixed during the fitting process.

Based on VLBA imaging observations of AGN jets, it is generally accepted that the jet exhibits a conical geometry with an opening angle $\theta_{\rm open}$ \citep{2007ApJ...668L..27K}. If the emission region fills the entire cross section of the jet, the radius of the region $R$ is related to its distance from the central engine $D$ by the relation $R=D\tan\theta_{\rm open}$, which is typically approximated as $R\approx 0.1D$ \citep{2009MNRAS.397..985G, 2012MNRAS.423..756P}. However, the variability timescale $t_{\rm var}$ can be used to constrain the radius $R$, following the relation $R\le ct_{\rm var}\delta/(1+z)$. Given a $t_{\rm var}$ of 0.36 days or less in the $\gamma$-ray band \citep{2021MNRAS.507.5602P}, the radius of the high-energy zone is estimated at $R_{\rm HE} \sim 10^{15}\ \rm cm$, while the radio band exhibits a $t_{\rm var}$ of weeks to months, corresponding to a radio zone radius of $R_{\rm radio} \sim 10^{17}\ \rm cm$.

Given the significant $\gamma$-ray emission in the high-energy zone, it is crucial to address the issue of $\gamma$-ray optical depth arising from $\gamma$--$\gamma$ absorption, where $\gamma$-ray photons interact with soft photons from both the emission region and the external environment. To avoid $\gamma$-ray absorption within the emission region, the nonthermal radiation must be strongly Doppler boosted. \citet{1995MNRAS.273..583D} proposed that the optical depth follows the relation $\tau_{\gamma\gamma} \propto \delta^{-(4+2\alpha)}$, where $\alpha$ is the spectral index, implying that a high $\delta$ effectively suppresses the optical depth. Given that the region is considered optically thin when $\tau_{\gamma\gamma} \leq 1$, \citet{2014RAA....14.1135F} derived a lower limit of $\delta \geq 2.61$ for the emission region of BL Lac to remain transparent to $\gamma$-rays. Therefore, the Doppler factor of the high-energy zone must satisfy this requirement. Moreover, it is also important to consider that the soft photons from the BLR can induce $\gamma$--$\gamma$ absorption of $\gamma$-ray photons. To avoid such absorption, it is necessary for the emission region to lie outside or near the outer edge of the BLR, especially during $\gamma$-ray flaring episodes, as discussed in \citet{2016ApJ...821..102B} and \citet{2021MNRAS.507.5602P}.

Overall, the $R_{\rm HE}$ and $D$ of the high-energy region, as well as the $R_{\rm radio}$ of the radio emission region, are left as free parameters within reasonable ranges. In contrast, the $D$ of the radio emission region is fixed to the relative separation $\Delta d$ obtained in Section~\ref{4.2}. The radii constrained by the variability timescales may be more compact, leading to $R<D \tan\theta_{\rm open}$. This implies that the size of the emission region could be smaller than the cross section of the jet.

\begin{figure*}[ht!]
\plotone{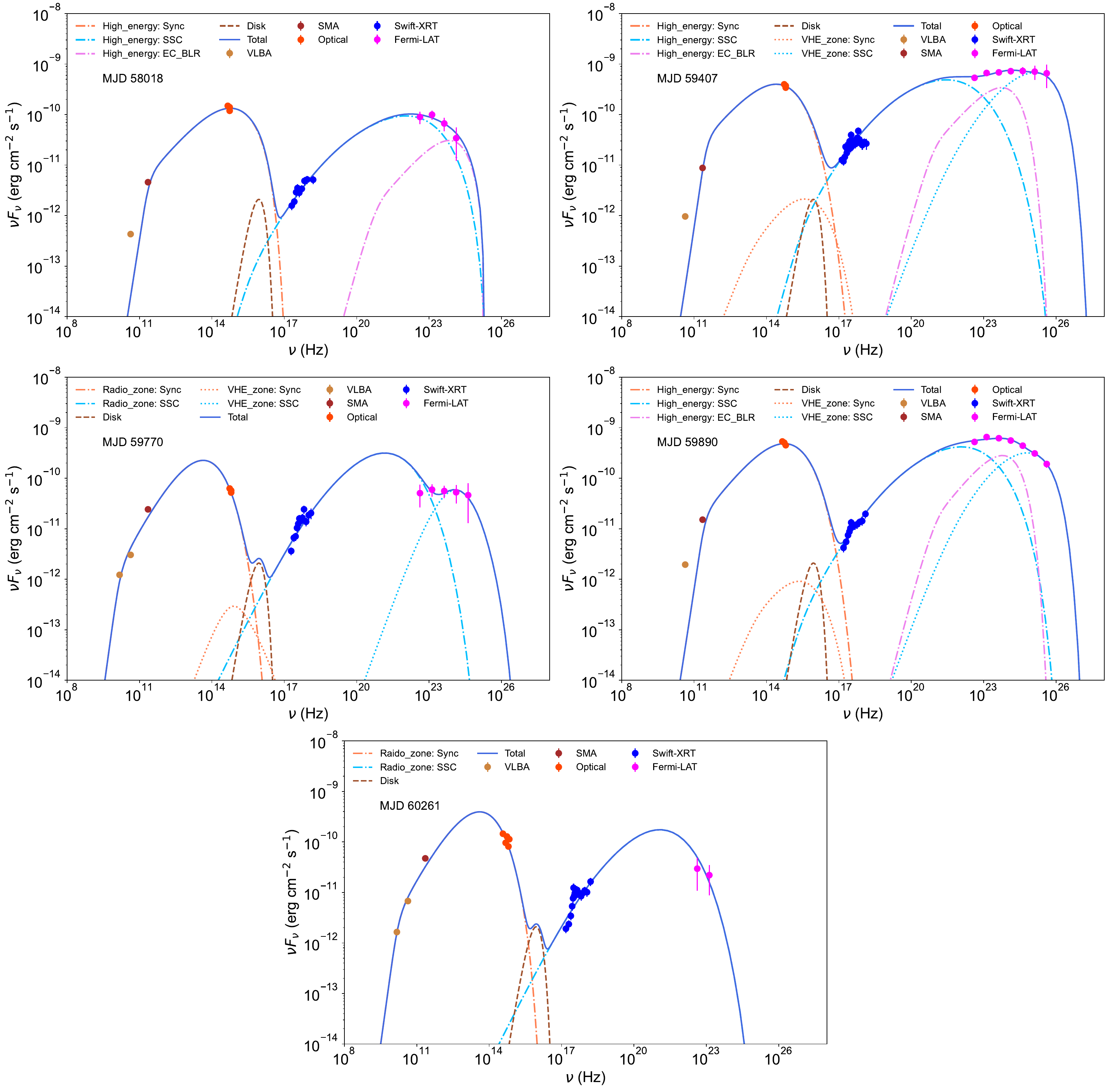}
\caption{Broad-band SEDs of BL Lac in the five selected epochs. Top row: the SED of the quiescent state on MJD 58018 and that of the first high-energy flare on MJD 59407, middle row: the SED of the first radio flare on MJD 59770 and that of the second high-energy flare on MJD 59890, bottom row: the SED of the second radio flare on MJD 60261, respectively.
\label{fig9}}
\end{figure*}

\subsection{Modeling Results and Discussion} \label{5.2}

\begin{deluxetable*}{lccccc}[ht!]
\tablewidth{0pt}
\tabletypesize{\small}
\tablecaption{Best-fit Parameters of Broad-band SEDs\label{tab5}}
\tablehead{
\colhead{High-energy} & \colhead{MJD 58018} & \colhead{MJD 59407} & \colhead{MJD 59770} & \colhead{MJD 59890} & \colhead{MJD 60261}
}
\startdata
$\gamma_{\rm min}$ & $57.76\pm5.21$ & $48.21\pm0.23$ & & $48.19\pm1.31$ & \\
$\gamma_{\rm max}$ & $(1.82\pm0.15)\times10^4$ & $(7.96\pm0.01)\times10^4$ & & $(9.29\pm0.02)\times10^4$ & \\
$\gamma_{\rm cut}$ & $(4.57\pm0.73)\times10^3$ & $(3.33\pm0.04)\times10^3$ & & $(4.04\pm0.04)\times10^3$ & \\
$p$ & $1.84\pm0.03$ & $1.84\pm0.07$ & & $1.74\pm0.02$ & \\
$R$ (cm) & $(5.71\pm0.10)\times10^{15}$ & $(6.23\pm0.12)\times10^{15}$ & & $(6.46\pm0.12)\times10^{15}$ & \\
$D$ (cm) & $4.54\times10^{16}$ & $5.06\times10^{16}$ & & $4.73\times10^{16}$ & \\
$B$ (G) & $(3.07\pm0.10)\times10^{-1}$ & $(1.64\pm0.01)\times10^{-1}$ & & $(2.02\pm0.08)\times10^{-1}$ & \\
$\Gamma$ & $14.54\pm0.55$ & $29.82\pm0.69$ & & $45.62\pm0.81$ & \\
$\theta$ ($^{\circ}$) & $1.40\pm0.02$ & $1.39\pm0.01$ & & $1.45\pm0.03$ & \\
$\rm log_{10}\nu_{\rm p}^{\rm sync}$ (Hz) & 14.78 & 14.40 & & 14.76 & \\
\hline
\sidehead{Radio}
\hline
$\gamma_{\rm min}$ & & & $1.97\pm0.01$ & & $1.68\pm0.09$\\
$\gamma_{\rm max}$ & & & $(3.89\pm0.01)\times10^4$ & & $(2.63\pm0.06)\times10^4$\\
$\gamma_{\rm cut}$ & & & $(2.89\pm0.10)\times10^3$ & & $(2.76\pm0.02)\times10^3$\\
$p$ & & & $1.07\pm0.02$ & & $1.10\pm0.03$\\
$R$ (cm) & & & $(2.25\pm0.12)\times10^{17}$ & & $(1.86\pm0.10)\times10^{17}$\\
$D$ (cm, fixed) & & & $4.50\times10^{19}$ & & $4.50\times10^{19}$\\
$B$ (G) & & & $(5.21\pm0.08)\times10^{-2}$ & & $(2.64\pm0.40)\times10^{-2}$\\
$\Gamma$ & & & $6.64\pm0.13$ & & $13.46\pm1.39$\\
$\theta$ ($^{\circ}$) & & & $5.68\pm0.07$ & & $1.72\pm0.04$\\
$\rm log_{10}\nu_{\rm p}^{\rm sync}$ (Hz) & & & 13.63 & & 13.58\\
\hline
\sidehead{VHE}
\hline
$\gamma_{\rm min}$ & & $(4.91\pm0.27)\times10^3$ & $(1.94\pm0.11)\times10^4$ & $(6.38\pm0.08)\times10^3$ & \\
$\gamma_{\rm max}$ & & $(1.33\pm0.07)\times10^6$ & $(7.35\pm1.13)\times10^5$ & $(2.72\pm0.25)\times10^6$ & \\
$\gamma_{\rm cut}$ & & $(4.81\pm0.20)\times10^4$ & $(7.63\pm0.78)\times10^4$ & $(3.99\pm0.03)\times10^4$ & \\
$p$ & & $1.21\pm0.03$ & $3.93\pm0.66$ & $1.22\pm0.01$ & \\
$R$ (cm, fixed) & & $1.00\times10^{15}$ & $1.00\times10^{15}$ & $1.00\times10^{15}$ & \\
$D$ (cm, fixed) & & $2.00\times10^{17}$ & $2.00\times10^{17}$ & $2.00\times10^{17}$ & \\
$B$ (G) & & $(5.25\pm0.21)\times10^{-3}$ & $(2.39\pm0.24)\times10^{-2}$ & $(6.52\pm0.06)\times10^{-3}$ & \\
$\Gamma$ & & $21.57\pm0.68$ & $10.06\pm0.83$ & $13.54\pm0.08$ & \\
$\theta$ ($^{\circ}$) & & $1.39\pm0.01$ & $5.68\pm0.07$ & $1.45\pm0.03$ & \\
\enddata
\tablecomments{The top panel lists the high-energy region parameters for MJDs 58018, 59407 and 59890. The middle panel lists the radio emission region parameters for MJDs 59770 and 60261. The bottom panel lists the VHE emission region parameters for MJDs 59407, 59770, and 59890.}
\end{deluxetable*}

Initially, the SEDs of the five epochs were modeled using a single-zone leptonic scenario. According to the discussion in Section~\ref{5.1}, for the quiescent state on MJD 58018 and the high-energy flaring states on MJDs 59407 and 59890, the radiation of BL Lac was mainly contributed by the high-energy region and the SEDs consisted of the synchrotron, SSC, and EC components, while for the radio flaring states on MJDs 59770 and 60261, the radiation was primarily produced by the radio emission region where the synchrotron and SSC components were included. However, in three epochs (MJDs 59407, 59770, and 59890), the emission in the GeV energy range could not be reproduced by the fitting. BL Lac is a TeV blazar and has been reported to have VHE radiation. This suggests the presence of a VHE emission region. Therefore, we added the VHE zone and adopted a two-zone leptonic model for the three epochs.  
Motivated by the extremely fast variability observed in the VHE band, we adopted a fixed radius of $R=1\times10^{15}\ \rm cm$ for the VHE zone as suggested by previous studies \citep{2019A&A...623A.175M, 2022MNRAS.513.4645S}. And we assumed that the region was placed at $D=2\times10^{17}\ \rm cm$, sited between the high-energy and radio zones. The VHE zone was set to be located beyond the BLR, which fulfills the $\gamma$-ray transparency requirement.
In order to limit the number of free parameters, the viewing angle $\theta$ of the VHE emission region was assumed to be the same as that of the high-energy or radio emission region. Being far from the BLR, the IC process in the VHE emission region involved only the SSC component.

The SED fitting curves corresponding to the five epochs are shown in Figure~\ref{fig9}, and the best-fit parameters are summarized in Table~\ref{tab5}. The fitting uncertainties for some free parameters are not provided in the table because they are negligibly small. One can see from Figure~\ref{fig9} that the overall SEDs generally agree well with the observations. For the SEDs on MJDs 58018, 59407, and 59890, the synchrotron radiation from the high-energy region dominates the radio-to-optical bands, while the SSC component accounts for the X-ray emission, and the soft $\gamma$-ray emission is produced by a combination of the SSC and EC processes. When BL Lac underwent a quiescent state on MJD 58018, the hard $\gamma $-ray emission was nearly absent (only upper limits are available). On MJDs 59407 and 59890, the hard $\gamma$-ray emission originates from the SSC process in the VHE emission region, whose synchrotron radiation contributes little to the overall SEDs. Notably, the VLBA observations on these three epochs show some excess flux compared to the fitting results, suggesting that some surplus radio emission of lower frequencies may originate from the downstream radio emission region. However, it should be noted that at these stages, the shock had not yet reached the radio emission zone to trigger flares, and the region remained relatively quiescent. On the other hand, in the SEDs of MJDs 59770 and 60261, the radiation over the radio-to-optical bands and the X-ray radiation are attributed to the synchrotron and SSC processes in the radio emission region, respectively. The soft $\gamma$-ray emission arises from the SSC process in the radio zone. As for the hard $\gamma$-ray radiation, the situation differs between the two epochs. For MJD 59770, although the $\gamma$-ray flux was about one order of magnitude lower than that of the $\gamma$-ray flaring states, hard $\gamma$-ray photons were still detected, suggesting that the SSC process of the VHE emission region still contributes to the hard $\gamma$-ray radiation. In contrast, for MJD 60261, almost no hard $\gamma$-ray photons were detected, implying that the VHE emission region makes no significant contribution at this time. 

\begin{figure}[ht!]
\begin{center}
    \includegraphics[angle=0,scale=0.3]{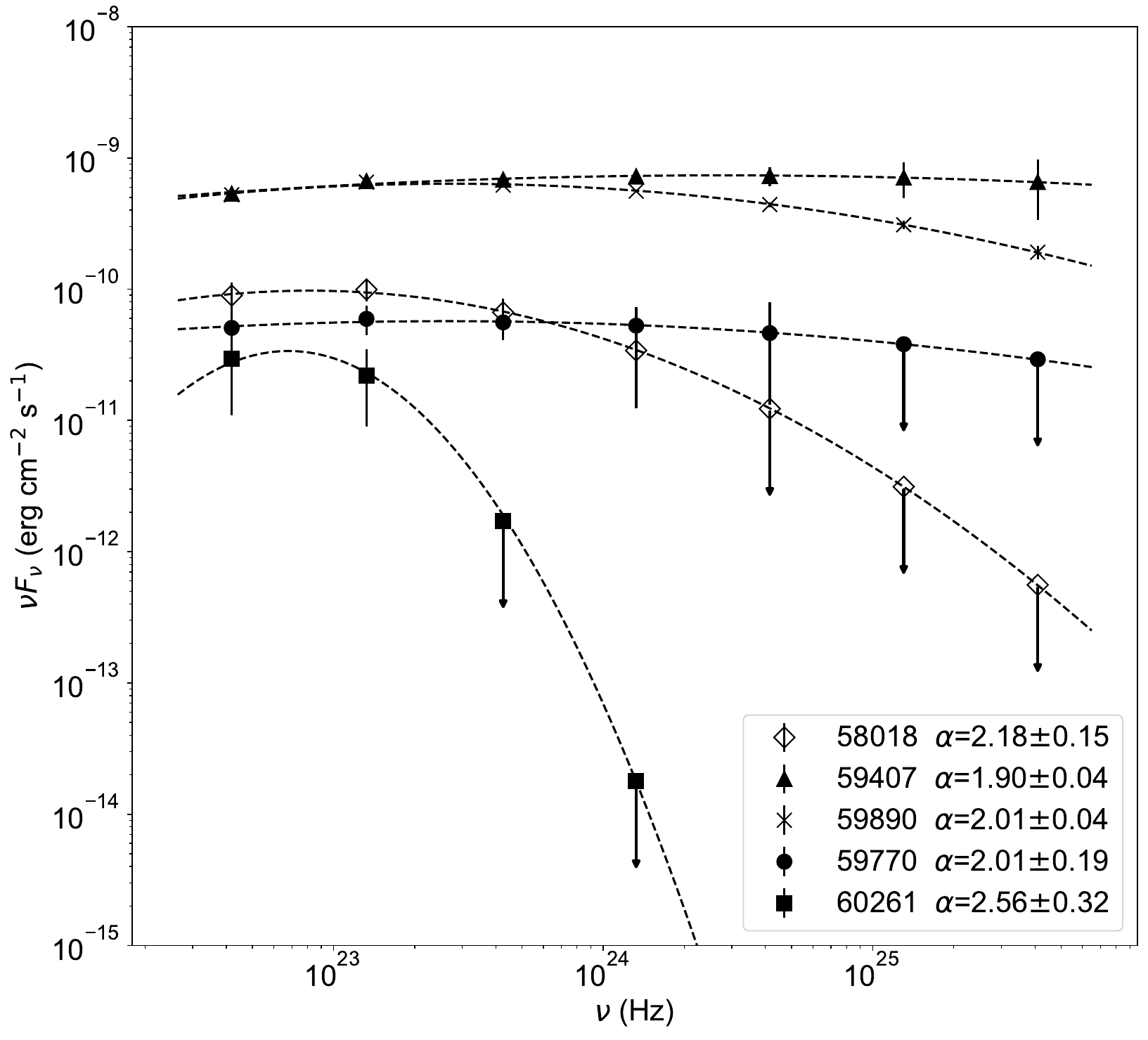}
\caption{$\gamma$-ray spectra of BL Lac in the five epochs. Downward arrows denote upper limits. Dates and corresponding photon indices $\alpha$ are indicated at the bottom right.
\label{fig10}}
\end{center}
\end{figure}

To further investigate the $\gamma$-ray emission in the five epochs and assess the possible existence of the VHE emission region, we employed a log-parabola model and derived the photon indices $\alpha$ when fitting the $\gamma$-ray spectra, as shown in Figure~\ref{fig10}. On the two dates associated with the $\gamma$-ray flares, MJDs 59407 and 59890, the values of $\alpha$ ($1.90\pm0.04$ and $2.01\pm0.04$, respectively) are relatively small, indicating hard $\gamma$-ray spectra. In contrast, during the $\gamma$-ray quiescent states, the $\alpha$ values on MJDs 58018 and 60261 ($2.18\pm0.15$ and $2.56\pm0.32$, respectively) are significantly larger, suggesting very soft $\gamma$-ray spectra at those epochs. However, the $\alpha$ of $2.01\pm0.19$ on MJD 59770 is comparable to that of the two flaring dates, implying that a hard $\gamma$-ray spectrum may still exist even during low $\gamma$-ray flux states. For a better comparison, we also obtained the average $\alpha$ from all observations in the LCR, which is $2.15\pm0.23$. In addition, \citet{2021MNRAS.507.5602P} reported an average $\alpha$ of 2.03 during MJD 59060--59260, which partly overlaps with the $\gamma$-ray flaring episode $\rm A_1$. Compared with these values, the $\gamma$-ray spectrum on MJD 59407 is clearly harder, those on MJDs 59770 and 59890 are moderately harder, while the spectra on MJDs 58018 and 60261 are distinctly softer. Taken together, a harder spectrum indicates higher-energy $\gamma$-ray photons, which in turn implies the existence of a VHE emission region.

The high-energy region is located around $5\times10^{16}\ \rm cm$ from the central engine, with a radial size at $\sim 6\times10^{15}\ \rm cm$, which meets the relation $R\approx 0.1D$. Compared with the out radius of the BLR, which is $1.9\times10^{16}\ \rm cm$, this emission region lies just outside the BLR, making it a reasonable inference that the seed photons could come from the BLR while avoiding $\gamma$--$\gamma$ absorption by the BLR photons. The high-energy region possesses a moderate bulk Lorentz factor ($\Gamma \sim 15$) in the quiescent state that increases to $\Gamma \gtrsim 30$ during the $\gamma$-ray flares, along with a small viewing angle of $\theta \sim 1.4^\circ$. The calculated values of the Doppler factor $\delta$ are 25.80, 39.15, and 39.11 for MJDs 58018, 59407, and 59890, respectively. A higher $\delta$ implies that the source appears to have more violent flares and significant variability.
At other epochs, higher values of $\delta$ have been reported, such as $\delta = 60$ on MJD 57188 \citep{2019A&A...623A.175M} and $\delta = 63$ on MJD 59117.4 \citep{2022MNRAS.513.4645S}.
It is worth noting that for the quiescent state on MJD 58018, the $\gamma_{\max}$ of the high-energy region is smaller by a factor of a few than those during the $\gamma$-ray flaring states, which implies a lower EED in the quiescent state. The internal shock model can account for the differences in parameters between the quiescent and high-energy flaring states \citep{2001MNRAS.325.1559S}. The shock is produced in this region, moves with a larger $\Gamma$, and accelerates the electrons within it. Furthermore, this region likely underwent two successive shock acceleration events, corresponding to the high-energy flares observed on MJDs 59407 and 59890, respectively.

The radius $R$ and the distance $D$ of the VHE region are fixed in our model, where $R<0.1D$. This represents a compact region within the jet, consistent with the rapid variability of the VHE radiation.
The VHE zone generally exhibits a higher EED than the high-energy region, resulting in its dominant contribution to the flux of the VHE bands and negligible contribution at lower-frequency bands. 
Compared to the high-energy region, where the minimum energy of the electrons is $\gamma_{\text{min}} \sim 50$ and the maximum is $\gamma_{\text{max}} \sim 10^4$, the VHE region has a significantly higher $\gamma_{\text{min}}$, reaching $10^3$ to $10^4$, and $\gamma_{\text{max}}$ can reach up to $10^6$.
The intensity of the magnetic field $B \lesssim 10^{-2}$ G in the VHE region is evidently lower than that in the high-energy region, indicating that the energy density of the electrons in the VHE zone outweighs that of the magnetic field. \citet{2022MNRAS.513.4645S} modeled an SED that includes MAGIC observations (MJD 57184.6--57199.7), and obtained a VHE emission region with a magnetic field intensity of $B = 3.32 \times 10^{-2}\ \mathrm{G}$ and an electron cutoff energy of $\gamma_{\text{cut}} = 1.39 \times 10^5$. In their work, the radiation from these electrons becomes dominant at energies above 30 GeV. 
The bulk Lorentz factor $\Gamma$ of the VHE zone is around 10--20, exhibiting modest enhancement during the $\gamma$-ray flaring states. Magnetic reconnection within the jet can accelerate and compress plasmoids, creating compact and fast-moving VHE emission regions. This scenario is referred to as the jets-in-a-jet or minijet model in the literature \citep{2009MNRAS.395L..29G, 2013MNRAS.431..355G}. Finally, the Doppler factors $\delta$ of the VHE and high-energy regions are high enough to guarantee the transparency to $\gamma$-rays.

The location of the radio emission region is fixed at $4.5 \times 10^{19}\ \mathrm{cm}$, and its radius obtained from the fit is $\sim 2 \times 10^{17}\ \mathrm{cm}$. 
Relative to its distance, the radio zone also appears slightly compact. VLBA imaging typically reveals that radio emission regions exhibit extended structures. Therefore, the radius derived in our model likely corresponds to the core of the radio zone, which is surrounded by more diffuse and low-density material.
Compared with the high-energy region, the radio emission region has a lower EED and a weaker magnetic field $B$, indicating that the electrons in this region produce more low-energy photons, corresponding to the enhanced radio flux observed. Moreover, in comparison with the high-energy region, the radio emission region has a smaller bulk Lorentz factor $\Gamma$ of around 10 and a larger viewing angle $\theta > 1.7^\circ$, implying a weaker Doppler boosting effect. This accounts for the relatively smoother variability in the radio bands, as discussed in Section~\ref{4.1}. Interestingly, between the epochs of MJDs 59770 and 60261, the bulk Lorentz factor $\Gamma$ and viewing angle $\theta$ of the radio emission region underwent notable changes. 
On MJD 60261, the region exhibited a larger $\Gamma = 13.46$ and a smaller $\theta = 1.72^\circ$ compared to the $\Gamma$ of 6.64 and $\theta$ of $5.68^\circ$ on MJD 59770.
This corresponds to the greater variation amplitude observed during episode $\rm B_2$ than that during episode $\rm B_1$. 
The variation of the $\Gamma$ can be attributed to shock acceleration, similar to the situation in the high-energy region. Furthermore, the correlation analysis of the variability corroborates the lag of the radio flares resulted from shock propagation. The change in viewing angle from $5.68^\circ$ to $1.72^\circ$ suggests a variation in the orientation of the jet relative to our line of sight, likely driven by jet precession or the helical motion of structures inside the jet.
Using the 15 GHz and $R$-band light curves of BL Lac from MJD 58450 to 59650, \citet{2024A&A...692A..48R} uncovered that the viewing angle of the emission region evolves over time.
Specifically, they proposed a geometric scenario in which the jet consists of a pair of emitting plasma filaments in a sort of curved double-helix rotating structure. The wiggling motion of this structure induces variations in the viewing angle and Doppler boosting effect (see Figure 5 therein). Such twisted plasma filaments may arise from Kelvin-Helmholtz or current-driven kink instabilities. This wiggling motion can also account for the minor changes in the $\theta$ of the high-energy region, and such changes are expected to be smaller due to the location of the region closer to the jet base.

In the following, we discuss the physical origins of the differences in the magnetic field $B$, the bulk Lorentz factor $\Gamma$, and the viewing angle $\theta$ between the three regions. First, within the framework of a uniform conical jet model with a constant $\Gamma$, the magnetic field $B$ is inversely proportional to the jet cross-sectional radius $R_{\rm j}$ ($B \propto R_{\rm j}^{-1}$) to ensure the conservation of magnetic energy \citep{2012MNRAS.423..756P}. However, our modeling results indicate that different emission regions are dominated by local physical processes and are in different physical states, which may cause the magnetic field to deviate from the aforementioned scaling law. In the high-energy and radio zones, shocks not only accelerate particles but also amplify the tangled magnetic field, while the magnitude of enhancement diminishing as the distance increases and the collision efficiency declines \citep{2001MNRAS.325.1559S}. In the VHE region, the magnetic reconnection dissipates magnetic energy into particle kinetic energy, and the electrons are accelerated to extremely high energies, but at the cost of a decrease in the local magnetic field strength $B$ \citep{2013MNRAS.431..355G}. The current-driven instabilities are likely to be the most relevant motivation to trigger the magnetic reconnection, and shocks seem to strengthen the instabilities according to our modeling results. Second, the average bulk Lorentz factors $\Gamma$ of the three emission regions follow a decreasing trend with distance, even though each region undergoes distinct local acceleration processes. This behavior is consistent with the scenario of the deceleration of the relativistic flow caused by radiative losses \citep{2003ApJ...594L..27G}, as also reflected in previous multi-zone SED modeling studies \citep{2019A&A...623A.175M, 2025ApJ...978...43M}. Finally, because of the spatial separation of the high-energy and radio zones along the jet, coupled with the rotational motion of the twisted double-helix structure, a discrepancy in their viewing angles $\theta$ is physically plausible. The $\theta$ of the VHE zone is fixed to be identical to that of the other emission region in the same epoch to limit the parameters in our model, so we do not discuss the physical origins of its variation in this work.

The synchrotron peak frequencies of the SEDs in the five epochs were calculated, and the results are presented in Table~\ref{tab5}. According to the classification introduced in Section~\ref{1}, BL Lac is classified as an IBL on MJDs 58018, 59407, and 59890, while it falls into the LBL category on MJDs 59770 and 60261. This confirms that BL Lac behaves as an IBL during quiescent and high-energy flaring states, as generally expected, and temporarily converts to an HBL \citep{2021MNRAS.507.5602P, 2025MNRAS.536.1251W}. In contrast, BL Lac tends to behave as an LBL during the radio flaring state.

For the flaring phase after MJD 55700, \citet{2013MNRAS.436.1530R} investigated the SED of BL Lac by selecting six epochs between MJD 55540 and 56040 during the $\gamma$-ray flaring state. They found that in two of these epochs, the IC peak frequency is located in the MeV range, while in the other four epochs it is in the GeV range. In our work, the IC peak frequency lies in the GeV band on MJDs 58018, 59407, and 59890 (corresponding to    quiescent and high-energy flaring states), whereas on MJDs 59770 and 60261 (corresponding to radio flaring states), it shifts to the MeV band. \citet{2016ApJ...816...53W} extracted multiwavelength data of six dates spanning MJD 56229--56304 during the radio flaring state to perform SED modeling of BL Lac. According to their model, the bulk Lorentz factor is $\Gamma = 6$, the viewing angle is $6^\circ$, and the electrons can be accelerated to energies up to $\gamma \sim 2\times10^{5}$. They infer a synchrotron peak frequency between $10^{13}$ and $10^{14}$ Hz and an IC peak frequency between $10^{20}$ and $10^{23}$ Hz, consistent with the double-peaked spectral shape of an LBL. Their fitting results generally agree well with our SED modeling for the radio flaring states on MJDs 59770 and 60261, except that they seem to derive a higher EED.

\section{Conclusion} \label{6}

In this study, we performed multicolor optical observations over 12 nights using the 85 cm telescope at Xinglong Station. Additionally, we collected multiwavelength observational data of BL Lac, covering radio, optical, X-ray, and $\gamma$-ray bands. A detailed investigation was conducted on the variability of this source on different timescales, as well as on its SED modeling. IDV was examined using the $F$-test and ANOVA, and the color behavior during IDV was analyzed. Time lags in IDV and LTV were explored utilizing three cross-correlation analysis methods: ICCF, \texttt{PyROA}, and ZDCF. Finally, the SEDs of the source were modeled adopting the \texttt{JetSeT} code, from which the physical parameters of the emission regions were derived, and flare mechanisms and spectral evolution were studied. The main conclusions of our study are as follows:

Among the 12 intraday light curves, IDV was detected on four nights, with the largest amplitude occurring in the $B$-band on MJD 59888. It was also found that higher-frequency bands exhibit larger amplitudes. Most of the IDVs follow the BWB trend and four spectral hysteresis loops were identified. Notably, both clockwise and counterclockwise loops were found simultaneously on MJD 59111 for the first time, reflecting the complex particle acceleration and cooling mechanisms in BL Lac. However, the time lag detection of the IDV did not yield any reasonable non-zero results.

We divided the optical to $\gamma$-ray light curves after MJD 58850 into three flaring episodes, and split the radio light curves into two flaring episodes. On MJD 60588.5, Fermi-LAT detected the highest $\gamma$-ray flux to date, reaching $(6.59 \pm 0.21) \times 10^{-6}\ \mathrm{phs\ cm^{-2}\ s^{-1}}$. The radio variability of BL Lac shows a superposition of a long-term trend and year-scale flares. A systematic cross-correlation analysis of the multiwavelength variations during this flaring period, for the first time, indicates that the time lags from the optical to $\gamma$-ray variations are approximately zero, while the average time lag between the radio and these higher-energy bands is about 370 days, suggesting the propagation of a shock within the jet. Based on this, there exist a high-energy region and a radio emission region in BL Lac, and the distance between them is estimated to be $4.50 \times 10^{19}\ \mathrm{cm}$, i.e., 14.58 pc.

Modeling of the SEDs in the five epochs, along with the analysis of the $\gamma$-ray spectra, suggests that in addition to the presence of the high-energy region and the radio emission region, in most cases BL Lac likely hosts a VHE emission region characterized by an extremely energetic EED, responsible for the production of VHE $\gamma$-ray photons. The high-energy region is located just outside the BLR, possessing a higher EED, a stronger magnetic field $B$, a larger bulk Lorentz factor $\Gamma$, and a smaller viewing angle $\theta$ than the radio emission region. The radio emission zone lies farthest from the central engine and has the most diffuse size. The synchrotron peak frequencies $\nu_{\rm p}^{\rm sync}$ in the five epochs demonstrate the spectral evolution of BL Lac: it behaves as an IBL when it is quiescent or flares in the optical to $\gamma$-ray bands, and converts into an LBL when it flares in the radio bands.

Our modeling results are based on the premise that jet emission is dominated by multiple discrete blobs subject to various acceleration mechanisms, which reflects the intricate physics within the jet. A uniform conical jet model may fail to fully reproduce the observations, especially when the source is in a flaring state, where local physical processes likely play a major role. In addition, alternative scenarios have been put forward to interpret the VHE emission. For example, \citet{2019A&A...623A.175M} proposed that the interaction between a compact blob embedded inside a larger one can reproduce the VHE observations. The energetic protons that interact in the jet can also form a population of energetic electrons \citep{2022MNRAS.513.4645S}. Therefore, the origin of the VHE emission in BL Lac is still a matter of debate. Future studies could explore multi-zone models that adhere to the uniform conical jet geometry and the magnetic field scaling law.

Although BL Lac is one of the most extensively studied blazars, it exhibited rich variability features during this prolonged flaring period, providing an opportunity to conduct a deep investigation of this source and gain further insights into the general properties of blazars. Notably, BL Lac underwent a new episode of $\gamma$-ray flaring after MJD 60500 (i.e., episode $\rm A_3$), highlighting the necessity for continued multiwavelength monitoring of the source.
 
\begin{acknowledgments}

This work has been supported by the Chinese National
Natural Science Foundation grant No. 12333001 and by the
National Key R\&D Program of China (2021YFA0718500 and 2025YFA1614101). This work made use of public Fermi-LAT data and data supplied by the UK Swift Science Data Centre at the University of Leicester. We acknowledge with thanks the observations from the AAVSO contributed by observers worldwide and used in this research. Data from ZTF, ASAS-SN, and KAIT have been used. The authors thank Mark Gurwell for sharing the SMA data with us. The SMA is a joint project between the Smithsonian Astrophysical Observatory and the Academia Sinica Institute of Astronomy and Astrophysics and is funded by the Smithsonian Institution and the Academia Sinica located near the summit of Maunakea in Hawaii. We recognize that Maunakea is a culturally important site for the indigenous Hawaiian people; we are privileged to study the cosmos from its summit. This study has made use of VLBA data from the VLBA-BU Blazar Monitoring Program (BEAM-ME and VLBA-BU-BLAZAR), funded by NASA through the Fermi Guest Investigator Program, and data from the MOJAVE database that is maintained by the MOJAVE team. 

\end{acknowledgments}

\facilities{Fermi, Swift(XRT), AAVSO, ASAS-SN, ZTF, KAIT, SMA, VLBA}

\software{Photutils \citep{2024zndo..10967176B}, Fermitools \citep{2019ascl.soft05011F}, HEASoft \citep{2014ascl.soft08004N}, XRT-CALDB \citep{2009A&A...494..775G}, BCES \citep{1996ApJ...470..706A, 2012Sci...338.1445N}, PyCCF \citep{1998PASP..110..660P, 2018ascl.soft05032S}, PyROA \citep{2021MNRAS.508.5449D}, ZDCF \citep{1997ASSL..218..163A, 2013arXiv1302.1508A}, pyPETaL \citep{2024ApJS..272...26S, 2024ascl.soft01004S}, JetSeT \citep{2009A&A...501..879T, 2011ApJ...739...66T, 2020ascl.soft09001T}}

\bibliography{reference}{}
\bibliographystyle{aasjournalv7}

\end{document}